\documentclass[a4paper,11pt]{article}
\pdfoutput=1 

\usepackage{jheppub} 
\usepackage{physics}
\usepackage{amsmath}
\usepackage{bbm}
\usepackage{xcolor}
\usepackage{mathtools}
\usepackage{tensor}
\usepackage{empheq}
\usepackage[makeroom]{cancel}

\newcommand\fft[2]{\frac{#1}{#2}}
\newcommand\ft[2]{{\textstyle\frac{#1}{#2}}}
\newcommand\nn{\nonumber}

\makeatletter
\newcommand*{\rom}[1]{\expandafter\@slowromancap\romannumeral #1@}
\makeatother

\preprint{LCTP-21-07}

\title{Subleading corrections to the $S^3$ free energy of necklace quiver theories dual to massive IIA}

\author{Junho Hong and James T. Liu}

\affiliation{Leinweber Center for Theoretical Physics, Randall Laboratory of Physics\\The University of Michigan, Ann Arbor, MI 48109-1040, USA }

\emailAdd{junhoh@umich.edu, jimliu@umich.edu}

\abstract{We investigate the $S^3$ free energy of $\mathcal N=3$ Chern-Simons-matter quiver gauge theories with gauge group U$(N)^r~(r\geq2)$ where the sum of Chern-Simons levels does not vanish, beyond the leading order in the large-$N$ limit. We take two different approaches to explore the sub-leading structures of the free energy. First we evaluate the matrix integral for the partition function in the 't~Hooft limit using a saddle point approximation. Second we use an ideal Fermi-gas model to compute the same partition function, but in the limit of fixed Chern-Simons levels. The resulting expressions for the free energy $F=-\log Z$ are then compared in the overlapping parameter regime. The Fermi-gas approach also hints at a universal $\fft16\log N$ correction to the free energy. Since the quiver gauge theories we consider are dual to massive Type~\rom{2}A theory, we expect the sub-leading correction of the planar free energy in the large 't Hooft parameter limit, which is one of our main results, to match higher-derivative corrections to the holographic dual free energy, which have not yet been fully investigated.}

\begin{document}
	\maketitle
	\flushbottom

\section{Introduction}

Ever since the advent of AdS/CFT, there have been tremendous progress in investigating and refining the duality.  Of particular interest are precision tests that can support the robustness of AdS/CFT beyond the leading order.  One example of such a test is the comparison of the free energy computed in the field theory with its holographic dual.  One immediate obstacle, however, is that the AdS/CFT correspondence is a strong-weak duality.  Thus the weak curvature limit that one needs to compute the free energy on the gravity side corresponds to a strong coupling limit that makes it hard to compute on the field theory side.

Supersymmetric localization \cite{Pestun:2007rz} has opened a way to overcome this issue. For superconformal field theories (SCFT), one can compute the partition function $Z$ precisely using supersymmetric localization that reduces the infinite-dimensional Euclidean path integral for the partition function into a matrix model. Since the result is valid even in the strong coupling limit, it can be compared with the holographic dual free energy $F=-\log Z$ in the weak curvature limit. This technical development motivates the calculation of supersymmetric partition functions of a wide class of SCFTs with holographic duals.  In addition to superconformal index computations, important examples include the $S^4$ partition function of $\mathcal N=4$ Super-Yang-Mills theory \cite{Pestun:2007rz} and the $S^3$ partition function of Chern-Simons-matter theories \cite{Kapustin:2009kz}. 

Of particular interest is the $S^3$ partition function of ABJM theory, namely $\mathcal N=6$ Chern-Simons-matter theory with gauge group U$(N)_k\times$U$(N)_{-k}$ and Chern-Simons levels $k$ and $-k$ \cite{Aharony:2008ug}. The $S^3$ free energy (and various Wilson loops) of ABJM theory was computed in the 't~Hooft limit --- the large-$N$ limit with fixed 't~Hooft coupling $\lambda=N/k$ --- using resolvent methods \cite{Marino:2009jd,Drukker:2010nc}, and in the M-theory limit by mapping to an ideal Fermi gas \cite{Marino:2011eh,Marino:2012az}. In the strong 't~Hooft coupling limit, the resulting free energy matches the holographic dual free energy at leading order \cite{Drukker:2010nc}, or equivalently the regularized on-shell action of Type~\rom{2}A string theory in an AdS$_4\times\mathbb{CP}^3$ background \cite{Balasubramanian:1999re,Emparan:1999pm}.

In the M-theory limit, where the large-$N$ limit is taken with fixed Chern-Simons level $k$, the leading order holographic dual free energy is given by the on-shell action of M-theory on AdS$_4\times S^7/\mathbb Z_k$, and scales like $N^\fft32$ as expected from the degrees of freedom on $N$ coincident M2-branes \cite{Klebanov:1996un}. Similar analyses have been done in more general ABJM-like $\mathcal N\geq2$ Chern-Simons-matter quiver gauge theories with gauge group U$(N)^r~(r\geq2)$ where the sum of Chern-Simons levels is zero \cite{Herzog:2010hf,Jafferis:2011zi}. The $S^3$ free energy was computed in the M-theory limit using the saddle point approximation of a matrix model from supersymmetric localization and matched with the holographic dual free energy at leading order.  Here the dual free energy is given as the regularized on-shell action of M-theory on AdS$_4\times X_7$ where $X_7$ is a tri-Sasaki Einstein manifold determined by matching with the dual field theory.

The large-$N$ comparison between the $S^3$ free energy of Chern-Simons-matter quiver gauge theories and their holographic duals has been developed to encompass sub-leading corrections as well. For the ABJM theory, in particular, the full genus expansion of the $S^3$ partition function in the 't~Hooft limit has been computed and written simply in terms of an Airy function \cite{Drukker:2011zy,Fuji:2011km}%
\footnote{Refer to \cite{Fuji:2011km,Hatsuda:2012dt} for some discussions on non-perturbative contributions to the free energy including worldsheet and D2 instanton corrections.}.
This Airy function result was further reproduced and justified in the M-theory limit for general ABJM-like $\mathcal N\geq2$ Chern-Simons-matter quiver gauge theories by rewriting the partition function as that of an ideal Fermi-gas \cite{Marino:2011eh,Marino:2012az}. Expanding the Airy function gives not just the leading $\mathcal O(N^{\fft32})$ term, but also a subleading $\mathcal O(N^{\fft12})$ term and a universal $\fft14\log N$ correction to the free energy.  The $\mathcal O(N^{\fft12})$ term is consistent with 8-derivative corrections to the holographic dual free energy \cite{Bergman:2009zh,Aharony:2009fc,Bhattacharyya:2012ye}, while the $\fft14\log N$ correction is reproduced from a one-loop determinant contribution to the holographic dual free energy \cite{Bhattacharyya:2012ye}. These observations have improved our understanding of the connection between ABJM-like theories and their holographic duals on AdS$_4\times X_7$ beyond the leading order.

Besides the ABJM-like cases, Gaiotto and Tomasiello (GT) have investigated another interesting class of SCFTs obtained from the ABJM theory by allowing for the sum of Chern-Simons levels to take a non-zero value that can be identified with the Romans mass in dual massive Type~\rom{2}A supergravity \cite{Gaiotto:2009mv,Gaiotto:2009yz}. The $S^3$ free energy of the GT theory has been computed in various ways: using resolvents in the 't~Hooft limit \cite{Suyama:2010hr,Suyama:2013fua}; using the saddle point approximation of a matrix model \cite{Jafferis:2011zi} and in the Fermi-gas approach in the large-$N$ limit with fixed Chern-Simons levels \cite{Marino:2011eh}. The leading order results from the various methods are consistent with each other in the overlapping parameter regime, and in particular scale like $N^\fft53$ in the large-$N$ limit with fixed Chern-Simons levels. This is in perfect agreement with the leading order holographic dual free energy given from the regularized on-shell action of massive Type~\rom{2}A supergravity on AdS$_4\times X_6$ \cite{Aharony:2010af,Petrini:2009ur,Lust:2009mb}, where the internal manifold $X_6$ depends on the detailed structures of the dual field theories. Note that massive Type~\rom{2}A theory with non-zero Romans mass does not have a strong coupling limit in the weak curvature regime, where it can be identified with M-theory \cite{Aharony:2010af}. Hence, for the GT theory, the large-$N$ limit with fixed Chern-Simons levels does not correspond to the M-theory limit, and in particular the theory is still dual to massive Type~\rom{2}A theory.

In contrast with that of ABJM-like theories, the sub-leading corrections to the $S^3$ free energy of GT theory have not yet been studied in detail. In this paper, we therefore investigate such corrections for general GT-like $\mathcal N=3$ Chern-Simons-matter quiver gauge theories with gauge group U$(N)^r~(r\geq2)$ where the sum of Chern-Simons levels does not vanish. The results will be a stepping stone towards a full exploration of GT-like theories and their holographic duals in massive Type~\rom{2}A supergravity.

In the remaining part of the introduction, we set up the problem and summarize our main results for the free energy of GT-like theories. In section \ref{sec:saddle}, we explore $r$-node quiver theories in the 't~Hooft limit using a saddle point approximation. In section \ref{sec:Fermi}, we focus on the two node case with $r=2$, and employ the ideal Fermi-gas model introduced and reviewed in \cite{Marino:2011eh,Marino:2016new}.  Finally, we contrast both approaches in section \ref{sec:discussion}, and discuss potential future directions.

\subsection{Setup}
We consider a class of $\mathcal N=3$ Chern-Simons-matter necklace quiver theories with $r$ nodes and the gauge group U$(N)^r$.  After localization, the partition function on $S^3$ can be written as a matrix model \cite{Kapustin:2009kz}
\begin{equation}
    Z=\fft{1}{(N!)^r}\int\left(\prod_{a=1}^r\prod_{i=1}^N\fft{d\lambda_{a,i}}{2\pi}\right)\prod_{a=1}^r\left[\fft{\prod_{i>j}\left(2\sinh\fft{\lambda_{a,i}-\lambda_{a,j}}{2}\right)^2}{\prod_{i,j}2\cosh\fft{\lambda_{a,i}-\lambda_{a+1,j}}{2}}\exp\left(\fft{i}{4\pi}\sum_{i=1}^Nk_a\lambda_{a,i}^2\right)\right],\label{eq:Zr}
\end{equation}
where $k_a$ is the Chern-Simons level for node $a$, and it is to be understood that $\lambda_{r+1,i}=\lambda_{1,i}$.  When the sum of the Chern-Simons levels, $k=\sum_ak_a$, vanishes the theory is ABJM-like, and the leading order free energy scales as $N^{3/2}$ in the large-$N$ limit with fixed $k_a$ \cite{Herzog:2010hf}.  However, when $k\ne0$, the theory is of Gaiotto-Tomasiello form where $k$ is dual to the Romans mass of massive IIA theory \cite{Gaiotto:2009mv}.  In this case, the free energy $F=-\log Z$ has been computed in \cite{Jafferis:2011zi} using the saddle point evaluation around the eigenvalue distribution in the large-$N$ limit with fixed $k_a$. The result exhibits $N^{5/3}$ scaling as\footnote{We modified (8.4) of \cite{Jafferis:2011zi} by a factor of $\fft{r}{2}$ as $F^\text{Here}=\fft{r}{2}F^\text{There}$.}
\begin{equation}
    F=-\log Z=\fft{3^\fft53r^\fft23\pi}{20}e^{-\fft{\pi i}6}k^{1/3}N^{5/3}.\label{eq:F:Jaffe}
\end{equation}

For the two-node Gaiotto-Tomasiello case \cite{Gaiotto:2009mv}, the above partition function (\ref{eq:Zr}) reduces to
\begin{equation}
	Z=\fft{1}{(N!)^2}\int\fft{d^N\mu}{(2\pi)^N}\fft{d^N\nu}{(2\pi)^N}\fft{\prod_{i>j}[2\sinh(\mu_{ij}/2)]^2[2\sinh(\nu_{ij}/2)]^2}{\prod_{i,j}[2\cosh((\mu_i-\nu_j)/2)]^2}\exp[\fft{i}{4\pi}\sum_{i=1}^N(k_1\mu_i^2+k_2\nu_i^2)],
	\label{eq:Z2}
\end{equation}
where $\mu_{ij}=\mu_i-\mu_j$ and $\nu_{ij}=\nu_i-\nu_j$. In this two-node case, the free energy (\ref{eq:F:Jaffe}) has been partially reproduced in \cite{Marino:2011eh} using an ideal Fermi-gas model in the large-$N$ limit with $\hbar\theta\ll1$ as
\begin{equation}
    F=-\log Z=\fft{3^\fft532^\fft23\pi}{20}e^{-\fft{\pi ii}6}k^{1/3}N^{5/3}\left(1+\fft{\hbar^2\theta^2}{4}\right)^\fft13,\label{eq:F:Marino}
\end{equation}
where $k=k_1+k_2$ and
\begin{equation}
	2\pi i\theta=-\fft{1}{k_1}-\fft{1}{k_2},\qquad \fft{4\pi}{\hbar}=\fft{1}{k_1}-\fft{1}{k_2}.\label{eq:theta:hbar}
\end{equation}
Note that (\ref{eq:F:Marino}) is consistent with (\ref{eq:F:Jaffe}) under the assumption $\hbar\theta\ll1$.

Our goal is to compute sub-leading corrections to the free energy (\ref{eq:F:Jaffe}) in the large-$N$ limit using both a saddle point analysis and (in the two-node case) an ideal Fermi-gas model. To keep track of sub-leading corrections, we must use an appropriate expansion parameter. The expansion parameters are different in the two approaches:
\begin{subequations}
\begin{align}
    \text{Saddle Point Analysis}&:~\text{large-}N\text{ with fixed }\fft{k_a}{N}~~\&~~\text{small-}\fft{|k|}{N},\label{expansion:saddle}\\
    \text{Ideal Fermi-gas Model}&:~\text{large-}N\text{ with fixed }k_a~~\&~~\text{small-}\hbar.\label{expansion:Fermi}
\end{align}\label{expansion}%
\end{subequations}
While the leading order free energy, (\ref{eq:F:Jaffe}), was captured by a saddle point analysis with fixed $k_a$ in \cite{Jafferis:2011zi}, we work instead with fixed $k_a/N$ as in (\ref{expansion:saddle}), which is more appropriate for the saddle point approach to keep track of sub-leading orders \cite{Mezei:2013gqa}. This choice corresponds to the 't~Hooft limit, and we focus on the genus-zero (planar) free energy which admits a natural expansion in powers of the inverse 't~Hooft coupling, $|k|/N$.  In contrast, working in the ideal Fermi-gas model requires taking the large-$N$ limit with fixed $k_a$.

\subsection{Summary of the results}
We approach the computation of the sphere partition function of GT theory through two methods.  The first is a saddle point analysis performed in the 't~Hooft limit, where we determine the first subleading correction to the planar free energy.  While we work in general with an $r$-node quiver, the result for the two-node GT model can be expressed as
\begin{empheq}[box=\fbox]{equation}
    F=N^2\left[\fft{3^{\fft53}2^{\fft23}\pi}{20}e^{-\fft{\pi i}6}\kappa^{1/3}+\fft{\pi i}{48}\kappa\left(1+3\left(\fft{k_1-k_2}{k_1+k_2}\right)^2\right)+\cdots\right]+o(N^2),
\label{eq:F:saddle:summary}
\end{empheq}
where $\kappa=(k_1+k_2)/N$ is the inverse 't~Hooft coupling.  The first sub-leading correction of the planar free energy in the large 't~Hooft coupling limit, namely the second term within the box bracket of (\ref{eq:F:saddle:summary}), is one of our main results that extends the leading order term, (\ref{eq:F:Jaffe}). The next term in the planar free energy is of $\mathcal O(\kappa^{4/3})$, but its computation is beyond the scope of the present work. We also leave a comment on higher genus terms $o(N^2)$ in subsection \ref{sec:saddle:beyond}. The general $r$-node free energy that generalizes (\ref{eq:F:saddle:summary}) is given in (\ref{eq:F:plannar:saddle}).

The second approach we use is the Fermi-gas picture that is applied at fixed Chern-Simons levels $k_a$.  We extend the leading order result for the two-node case, (\ref{eq:F:Marino}), to the next order in the $\hbar$ expansion
\begin{empheq}[box=\fbox]{equation}
\begin{split}
    F&=\fft{1}{\pi\hbar\theta^3}\biggl[\fft15\left(\fft{3\pi\hbar\theta^2}{2}N\right)^\fft53\left(1-\fft{\hbar^2\theta^2}{12}+\fft{\hbar^4\theta^4}{72}+\mathcal O(\hbar^6)\right)\\
    &\kern4em+\fft{1}{12}\left(\fft{3\pi\hbar\theta^2}{2}N\right)\left(1-\fft{\hbar^2\theta^2}{3}+\fft{\hbar^4\theta^4}{12}+\mathcal O(\hbar^6)\right)\\
    &\kern4em-\fft{1}{12}\left(\fft{3\pi\hbar\theta^2}{2}N\right)^\fft23\left(1-\fft{\hbar^2\theta^2}{12}+\mathcal O(\hbar^4)\right)\\
    &\kern4em+\fft{1}{16}\left(\fft{3\pi\hbar\theta^2}{2}N\right)^\fft13\left(1-\fft{\hbar^2\theta^2(3-16\theta)}{12}+\mathcal O(\hbar^4)\right)\biggr]\\
    &\quad+\fft16\log N+\mathcal O(N^0).
\label{eq:F:Fermi:summary}
\end{split}
\end{empheq}
(Note that the $\mathcal O(\hbar^4)$ are obtained numerically.)  Several remarks are in order at this time:
\begin{itemize}
    \item The sub-leading correction of the planar free energy in the large 't~Hooft parameter (small $|\kappa|$) expansion, namely the second term in (\ref{eq:F:saddle:summary}), is expected to come from higher derivative corrections in the supergravity dual. Recent developments in \cite{Bobev:2020egg} could be useful to confirm this holographic relation.
    \item For the two node case, we can compare the two results, (\ref{eq:F:saddle:summary}) and (\ref{eq:F:Fermi:summary}), at least in the overlapping parameter regime where both expansions (\ref{expansion}) are valid. Interestingly enough, we find that the explicitly written terms in the small $|\kappa|$ expansion, (\ref{eq:F:saddle:summary}), matches the first two lines in the fixed $k_a$ expansion, (\ref{eq:F:Fermi:summary}), in the overlapping parameter regime. Refer to section \ref{sec:discussion} for details and comments about this match.
    \item The expansion (\ref{eq:F:Fermi:summary}) has a universal $\fft16\log N$ correction, which corresponds to a one-loop quantum correction in the dual supergravity.  This was not seen in the saddle point expansion since there we worked exclusively in the planar limit.
\end{itemize}
Additional discussion on these points can be found in section \ref{sec:discussion}.

\section{Saddle point analysis and the planar free energy}\label{sec:saddle}

The leading order expression for the free energy, (\ref{eq:F:Jaffe}), was obtained in the large-$N$ limit at fixed $k$.  However, it can be rewritten in a form suggestive of a 't~Hooft expansion
\begin{equation}
    F=N^2\left(\fft{3^\fft53r^\fft23\pi}{20}e^{-\fft{\pi i}6}\kappa^{1/3}+\cdots\right)+o(N^2),
\label{eq:FtHooft}
\end{equation}
where $\kappa\equiv k/N=\sum_ak_a/N$ is the \emph{inverse} `t~Hooft parameter.  The planar free energy is expected to receive corrections proportional to higher powers of $\kappa^{1/3}$, and they can be investigated using a saddle point analysis.

It is worth keeping in mind that the analysis that we perform involves two limits.  The first is the planar limit in the large-$N$ expansion, where we keep only the genus zero contribution to the free energy.  This allows us to directly replace discrete sums over eigenvalues by integrals over corresponding densities without the need for Euler-Maclaurin corrections, and moreover allows us to work entirely at the level of a classical effective action.  Even so, the saddle point equation does not appear to admit a simple solution except in the large 't~Hooft parameter limit.  We thus perform an additional expansion in small $\kappa$ and in this way obtain the $\mathcal O(\kappa)$ contribution to (\ref{eq:FtHooft}).

\subsection{The effective action in the \texorpdfstring{large-$N$}{large-N} limit}
\label{sec:saddle:setup}

We start by writing the $S^3$ partition function (\ref{eq:Zr}) in terms of an effective action
\begin{equation}
	Z(N,k_a)=\fft{1}{(N!)^r}\int\left(\prod_{a=1}^r\prod_{i=1}^N\fft{d\lambda_{a,i}}{2\pi}\right)\,e^{N^2S_{\mathrm{eff}}(\lambda_a;k_a)},\label{eq:Zr:Seff}
\end{equation}
where
\begin{equation}
\begin{split}
	N^2S_{\mathrm{eff}}(\lambda_a;k_a)&=\sum_{a=1}^r\biggl[\fft{i}{4\pi}\sum_{i=1}^Nk_a\lambda_{a,i}^2+2\sum_{i>j}^N\log(2\sinh\fft{\lambda_{a,i}-\lambda_{a,j}}{2})\\
	&\kern3em~-\sum_{i,j=1}^N\log(2\cosh\fft{\lambda_{a,i}-\lambda_{a+1,j}}{2})\biggr].
    \label{eq:Seff}
\end{split}
\end{equation}
The saddle point equations are obtained by demanding $\partial S_{\mathrm{eff}}/\partial\lambda_{a,i}=0$, and take the form
\begin{equation}
	0=\fft{ik_a}{2\pi}\lambda_{a,i}+\sum_{j=1\,(\neq i)}^N\coth\fft{\lambda_{a,i}-\lambda_{a,j}}{2}-\fft12\sum_{j=1}^N\left(\tanh\fft{\lambda_{a,i}-\lambda_{a+1,j}}{2}+\tanh\fft{\lambda_{a,i}-\lambda_{a-1,j}}{2}\right).\label{eq:saddle}
\end{equation}
In principle, the planar free energy is obtained by solving the saddle point equations and then inserting the solution into the effective action.  However, in practice it is infeasible to directly solve this set of $Nr$ coupled non-linear equations for the eigenvalues $\lambda_{a,i}$ in any realistic scenario.

We thus proceed by working in the large-$N$ limit where we replace the discrete eigenvalues $\lambda_{a,i}$ by continuous eigenvalue distributions $\lambda_a:I_a\subseteq\mathbb R\to\mathbb C$ according to
\begin{equation}
    \lambda_a(x^{(a)}(i))=\lambda_{a,i},\label{saddle:ansatz}
\end{equation}
where we have introduced a continuous function $x^{(a)}:[1,N]\to I_a\subseteq\mathbb R$ that maps discrete indices $1,\cdots,N$ to a subset of real line $I_a$. Without loss of generality, we assume $x^{(a)}$ is an increasing function as $x^{(a)}(i)<x^{(a)}(j)$ for $i<j$. 

In contrast with fixed $k_a$ expansions, we take the large-$N$ limit while holding the inverse 't~Hooft parameters
\begin{equation}
    \kappa_a\equiv\fft{k_a}{N},\qquad\kappa\equiv\sum_{a=1}^r\kappa_a,\label{eq:kappa}
\end{equation}
fixed.  In this limit, the effective action $S_\text{eff}$ is consistently of $\mathcal O(N^0)$ and represents the planar contribution to the free energy.  Working only at this order, we now take the continuum limit of (\ref{eq:Seff}) while discarding any contributions of $\mathcal O(1/N)$ or higher.  The result is
\begin{equation}
\begin{split}
	S_\text{eff}(\lambda_a,\rho_a;\kappa_a)&=\fft{i}{4\pi}\sum_{a=1}^r\kappa_a\int_{I_a}dx\,\rho_a(x)\lambda_a(x)^2\\
	&\quad+\sum_{a=1}^r\int_{I_a}dx\,\rho_a(x)\left[\int_{L[I_a]}^xdx'\,\rho_a(x')\log(2\sinh\fft{\lambda_a(x)-\lambda_a(x')}{2})\right.\\
	&\kern9em~\left.+\int_x^{R[I_a]}dx'\,\rho_a(x')\log(2\sinh\fft{\lambda_a(x')-\lambda_a(x)}{2})\right]\\
	&\quad-\sum_{a=1}^r\int_{I_a}dx\,\rho_a(x)\int_{I_{a+1}}dx'\,\rho_{a+1}(x')\log(2\cosh\fft{\lambda_a(x)-\lambda_{a+1}(x')}{2})\\
	&\quad+o(N^0),\label{eq:Seff:continuum}
\end{split}
\end{equation}
where $L[I_a]$ and $R[I_a]$ stand for the left and the right end of the domain $I_a$ respectively. Here we have also introduced the eigenvalue density $\rho_a:I_a\to\mathbb R$ as
\begin{equation}
	di=(N-1)\rho_a(x)dx\qquad\to\qquad \int_{I_a}dx\,\rho_a(x)=1.\label{eq:rho}
\end{equation}
Note that the effective action (\ref{eq:Seff:continuum}) is now a functional of $\lambda_a(x)$ and $\rho_a(x)$. We also replaced the Chern-Simons level $k_a$ in the last argument of (\ref{eq:Seff}) with the inverse 't~Hooft parameter $\kappa_a$ in (\ref{eq:Seff:continuum}) to emphasize that the latter is to be held finite in the large-$N$ limit.

Note that the vector multiplet contribution in (\ref{eq:Seff:continuum}) is split into two contributions depending on whether $x<x'$ or $x>x'$ in order to preserve the ordering $i>j$ in the discrete sum of (\ref{eq:Seff}).  The integrand is actually logarithmically divergent for $x=x'$, but such a divergence can be integrated.  The divergence along the $x=x'$ `diagonal' will lead to a $(1/N)\log N$ correction to the effective action, but this is unimportant in the planar limit.

\subsection{Expanding in the large 't~Hooft parameter limit}

While the continuum action (\ref{eq:Seff:continuum}) fully captures the genus zero contribution, it remains difficult to work with since it is non-local.  An important simplification occurs by noting that the long range forces between the $\log\sinh z$ and $\log\cosh z$ terms cancel.  This allows the action to be written in a quasi-local form that can be made explicit by expanding in the small $\kappa$ limit.

As noted both numerically and analytically, in the limit $\kappa\to0$, the eigenvalues condense on a single cut in the complex plane \cite{Jafferis:2011zi}.  Translating the fixed $k_a$ solution of \cite{Jafferis:2011zi} to the present fixed $\kappa_a$ case suggests that we parametrize the continuum eigenvalues as
\begin{equation}
    \lambda_a(x)=|\kappa|^{-\alpha}(x+iy(x))+|\kappa|^\alpha z_a(x)+\mathcal O(|\kappa|^{2\alpha}),\qquad x\in(-x_*,x_*),
\label{saddle:ansatz:smallk}
\end{equation}
where $x$ is a real variable and $\pm x_*$ are the endpoints of the cut.  While the saddle point solution requires $\alpha=1/3$, we keep $\alpha$ explicit here in order to keep track of the perturbative order in the small $\kappa$ expansion.  The leading order cut in the complex plane is parametrized by the real function $y(x)$, and is identical for all nodes in the quiver.  The first subleading correction is no longer universal, and is parametrized by the set of complex functions $z_a(x)$.

In addition, we introduce the eigenvalue density
\begin{equation}
    \rho_a(x)=\rho(x)\quad\mbox{where}\quad\rho(x)=\rho_0(x)+|\kappa|^{2\alpha}\rho_1(x)+\mathcal O(|\kappa|^{3\alpha}),\qquad x\in(-x_*,x_*).
\label{eq:rho:smallk}
\end{equation}
Note that we take all nodes to have the same eigenvalue density.  This is clearly valid for the leading order $\rho_0(x)$, but need not be the case at subleading order.  However, by making use of the freedom of perturbing the eigenvalues by $z_a(x)$, it is possible to make the subleading correction $\rho_1(x)$ independent of the node, and we choose to do so for simplicity.  The normalization of the eigenvalue density is then given in terms of $\rho_0(x)$ and $\rho_1(x)$ as
\begin{equation}
    \int_{-x*}^{x_*}dx\,\rho_0(x)=1,\qquad \int_{-x*}^{x_*}dx\,\rho_1(x)=0.
\label{normalization}
\end{equation}
It should be noted that the ans\"atze (\ref{saddle:ansatz:smallk}) and (\ref{eq:rho:smallk}) for the subleading corrections $z_a(x)$ and $\rho_1(x)$ assumes a perturbative expansion in $|\kappa|^\alpha$ that skips the first order so that the first subleading terms are of $\mathcal O(|\kappa|^{2\alpha})$ compared to the leading order.  This is not necessarily the most general possibility but nevertheless leads to a consistent expansion, as can be verified \textit{a postiori}.

We now substitute the expansion of $\lambda_a(x)$ and $\rho_a(x)$ into the effective action (\ref{eq:Seff:continuum}) and rewrite the $x'$ integrals by introducing a new integration variable $u=|\kappa|^{-\alpha}(x-x')$.  The effective action can then be expanded in powers of $|\kappa|^\alpha$ as
\begin{equation}
    S_\text{eff}(\lambda_a,\rho_a;\kappa_a)=\sum_{n=1}^\infty|\kappa|^{n\alpha}S_n(y,z_a,\rho_0,\rho_1,\ldots;\kappa_a)+o(N^0),
\label{eq:Seff:continuum:smallk}
\end{equation}
where the ellipsis denotes higher order terms in the ans\"atze that we have not written out explicitly.  For the first subleading correction, we need the first three terms in the expansion of $S_{\mathrm{eff}}$
\begin{subequations}
\begin{align}
	S_1&=\fft{i}{4\pi}\fft{\kappa}{|\kappa|}\int_{-x_\ast}^{x_\ast}dx\,\rho_0(x)(x+iy(x))^2\nn\\
	&\quad+r\int_{-x_\ast}^{x_\ast}dx\,\rho_0(x)^2\int_{-|\kappa|^{-\alpha}(x_\ast-x)}^{|\kappa|^{-\alpha}(x_\ast+x)}du\,\log\tanh\fft{(1+iy'(x))|u|}{2},\label{eq:Seff:1}\\
	S_2&=r\int_{-x_\ast}^{x_\ast}dx\,\rho_0(x)\int_{-|\kappa|^{-\alpha}(x_\ast-x)}^{|\kappa|^{-\alpha}(x_\ast+x)}du\,\left[-\rho_0'(x)u\log\tanh\fft{(1+iy'(x))|u|}{2}\right.\nn\\
	&\kern3em\left.+\rho_0(x)\left(\coth\fft{(1+iy'(x))u}{2}-\tanh\fft{(1+iy'(x))u}{2}\right)(-\fft{i}{4}y''(x)u^2)\right],\label{eq:Seff:2}\\
	S_3&=\fft{i}{4\pi}\fft{\kappa}{|\kappa|}\int_{-x_\ast}^{x_\ast}dx\,\rho_1(x)(x+iy(x))^2+\fft{i}{2\pi}\sum_{a=1}^r\fft{\kappa_a}{|\kappa|}\int_{-x_\ast}^{x_\ast}dx\,\rho_0(x)(x+iy(x))z_a(x)\nn\\
	&\quad+\sum_{a=1}^r\int_{-x_\ast}^{x_\ast}dx\,\rho_0(x)\int_{-|\kappa|^{-\alpha}(x_\ast-x)}^{|\kappa|^{-\alpha}(x_\ast+x)}du\,\left[\rho_0(x)\coth\fft{(1+iy'(x))u}{2}\left(\fft{i}{12}y'''(x)u^3+\fft{z_a'(x)}{2}u\right)\right.\nn\\
	&\kern3em+\fft12\rho_0(x)\csch^2\fft{(1+iy'(x))u}{2}\fft{1}{16}y''(x)^2u^4+\rho_0'(x)\coth\fft{(1+iy'(x))u}{2}\fft{i}{4}y''(x)u^3\nn\\
	&\kern3em\left.+(2\rho_1(x)+\fft12\rho_0''(x)u^2)\log\sinh\fft{(1+iy'(x))|u|}{2}\right]\nn\\
	&\quad-\sum_{a=1}^r\int_{-x_\ast}^{x_\ast}dx\,\rho_0(x)\int_{-|\kappa|^{-\alpha}(x_\ast-x)}^{|\kappa|^{-\alpha}(x_\ast+x)}du\,\left[\rho_0(x)\tanh\fft{(1+iy'(x))u}{2}\left(\fft{i}{12}y'''(x)u^3+\fft{z_a'(x)+z_{a+1}'(x)}{4}u\right)\right.\nn\\
	&\kern3em+\fft12\rho_0(x)\sech^2\fft{(1+iy'(x))u}{2}\left(-\fft{1}{16}y''(x)^2u^4+\fft14(z_a(x)-z_{a+1}(x))^2\right)\nn\\
	&\kern3em\left.+\rho_0'(x)\tanh\fft{(1+iy'(x))u}{2}\fft{i}{4}y''(x)u^3+(2\rho_1(x)+\fft12\rho_0''(x)u^2)\log\cosh\fft{(1+iy'(x))u}{2}\right].\label{eq:Seff:3}
\end{align}
\end{subequations}
These expressions can be viewed as the perturbative expansion of the effective action as a function of the leading order fields $y(x)$ and $\rho_0(x)$ and the subleading fields $z_a(x)$ and $\rho_1(x)$.

As written, the expansion of $S_{\mathrm{eff}}$ still involves double integrals.  However, each of the $S_n$ is local in the fields as they have been expanded around $x$.  This can be seen more explicitly by performing the $u$ integral.  This is not entirely straightforward because the endpoints of the integral are inherited from the endpoints of the one-cut solution.  Note, however, that the limits $\pm|\kappa|^{-\alpha}(x_*\pm x)$ approach $\pm\infty$ for small $\kappa$, expect in a small region around the endpoints, $x=\pm x_*+\mathcal O(|\kappa|^\alpha)$.  This suggests that we rewrite the $u$ integral as an improper integral plus correction terms
\begin{equation}
    \int_{-|\kappa|^{-\alpha}(x_\ast-x)}^{|\kappa|^{-\alpha}(x_\ast+x)}du=\int_\infty^\infty du-\int_{|\kappa|^{-\alpha}(x_\ast+x)}^\infty du-\int_{-\infty}^{-|\kappa|^{-\alpha}(x_\ast-x)}du.
\end{equation}
The correction terms are exponentially suppressed except at the endpoints.

For $S_1$, the $u$ integral can be evaluated explicitly, with the result \cite{Jafferis:2011zi}
\begin{equation}
	S_1=\fft{i}{4\pi}\fft{\kappa}{|\kappa|}\int_{-x_\ast}^{x_\ast}dx\,\rho_0(x)(x+iy(x))^2-\fft{r\pi^2}{2}\int_{-x_\ast}^{x_\ast}dx\,\fft{\rho_0(x)^2}{1+iy'(x)}+S_1^\text{end},\label{eq:Seff:1:withend}
\end{equation}
where the endpoint correction is given by
\begin{equation}
\begin{split}
	S_1^\text{end}&\equiv-r\int_{-x_\ast}^{x_\ast}dx\,\fft{\rho_0(x)^2}{1+iy'(x)}\left[\log\tanh\fft{(1+iy'(x))u}{2}\log(1+\tanh\fft{(1+iy'(x))u}{2})\right.\\
	&\kern3em\left.+\text{Li}_2(1-\tanh\fft{(1+iy'(x))u}{2})+\text{Li}_2(-\tanh\fft{(1+iy'(x))u}{2})\right]_{|\kappa|^{-\alpha}(x_\ast+x)}^\infty\\
	&\quad-r\int_{-x_\ast}^{x_\ast}dx\,\fft{\rho_0(x)^2}{1+iy'(x)}\left[\log\tanh\fft{(1+iy'(x))u}{2}\log(1+\tanh\fft{(1+iy'(x))u}{2})\right.\\
	&\kern3em\left.+\text{Li}_2(1-\tanh\fft{(1+iy'(x))u}{2})+\text{Li}_2(-\tanh\fft{(1+iy'(x))u}{2})\right]_{|\kappa|^{-\alpha}(x_\ast-x)}^\infty.\label{eq:Seff:1:end}
\end{split}
\end{equation}
This is exponentially suppressed as $\mathcal O(e^{-|\kappa|^{-\alpha}})$ except at the endpoints.

For $S_2$, the $u$ integral vanishes when extended to $\pm\infty$ since the integrand is an odd function of $u$.  However, an endpoint correction remains since the true integration range is not actually symmetrical.  We thus have
\begin{equation}
    S_2=S_2^{\mathrm{end}},
\label{eq:S2sub}
\end{equation}
where the endpoint correction is again exponentially suppressed except at the endpoints.  The vanishing of $S_2$ away from the end points is the reason that the perturbative corrections to $\lambda_a(x)$ and $\rho_a(x)$ skip an order in the $|\kappa|^\alpha$ expansion.

Finally, performing the $u$ integral in $S_3$ gives the result
\begin{equation}
\begin{split}
	S_3&=\fft{i}{4\pi}\int_{-x_\ast}^{x_\ast}dx\,\left[\fft{\kappa}{|\kappa|}\rho_1(x)(x+iy(x))^2+2\rho_0(x)(x+iy(x))\sum_{a=1}^r\fft{\kappa_a}{|\kappa|}z_a(x)\right]\\
	&\quad+\int_{-x_\ast}^{x_\ast}dx\,\rho_0(x)^2\left[-\fft{\sum_{a=1}^r(z_a(x)-z_{a+1}(x))^2}{2(1+iy'(x))}+\fft{\pi^2\sum_{a=1}^rz_a'(x)}{2(1+iy'(x))^2}\right.\\
	&\kern9em~\left.+\fft{r\pi^4i y'''(x)}{24(1+iy'(x))^4}+\fft{r\pi^4y''(x)^2}{8(1+iy'(x))^5}\right]\\
	&\quad+r\int_{-x_\ast}^{x_\ast}dx\,\rho_0(x)\left[\fft{\pi^4i\rho_0'(x)y''(x)}{8(1+iy'(x))^4}-\fft{\pi^4\rho_0''(x)}{24(1+iy'(x))^3}-\fft{\pi^2\rho_1(x)}{1+iy'(x)}\right]+S_3^\text{end},\label{eq:Seff:3:simple}
\end{split}
\end{equation}
where $S_3^\text{end}$ is the correction term that is exponentially suppressed away from the endpoints.

\subsection{Obtaining the saddle point solution}\label{sec:saddle:solution}

The leading order effective action (\ref{eq:Seff:1:withend}) was obtained in \cite{Jafferis:2011zi}, and depends on the leading order fields $\rho_0(x)$ and $y(x)$.  By varying with respect to these fields, we can obtain the leading order saddle point equations
\begin{subequations}
\begin{align}
	0&=\fft{i}{4\pi}\fft{\kappa}{|\kappa|}(x+iy(x))^2-\fft{r\pi^2\rho_0(x)}{1+iy'(x)}+\Lambda_1+\cancel{\fft{\delta S_1^\text{end}}{\delta\rho_0(x)}},\label{eq:saddle:1:rho0}\\
	0&=-\fft{1}{2\pi}\fft{\kappa}{|\kappa|}\rho_0(x)(x+iy(x))-\fft{r\pi^2i}{2}\fft{d}{dx}\fft{\rho_0(x)^2}{(1+iy'(x))^2}+\cancel{\fft{\delta S_1^\text{end}}{\delta y(x)}}.\label{eq:saddle:1:y}
\end{align}\label{eq:saddle:1}%
\end{subequations}
Here we have introduced a Lagrange multiplier $\Lambda_1$ to enforce the normalization of $\rho_0$.

Away from the endpoints, we can ignore the functional derivatives of $S_1^\text{end}$ since they are exponentially suppressed as $\mathcal O(e^{-|\kappa|^{-\alpha}})$ in the small-$|\kappa|$ expansion.  We also note that the second equation, (\ref{eq:saddle:1:y}), is in fact the $x$-derivative of the first equation, (\ref{eq:saddle:1:rho0}).  As a result, we only need to solve the first equation.  To do so, we assume a constant slope solution, $y(x)=mx$, with $m\in\mathbb R$. Demanding $\rho_0(x)$ to be normalized and a real function of $x$ then leads to the leading order saddle point solution \cite{Jafferis:2011zi}
\begin{empheq}[box=\fbox]{align}
	y(x)=mx,\qquad
	\rho_0(x)=\fft3{4x_*}\left(1-\fft{x^2}{x_*^2}\right),
\label{sol:saddle:1}
\end{empheq}
where
\begin{equation}
    \left(\fft{1+im}{\sqrt{1+m^2}}\right)^3=i\fft{|\kappa|}{\kappa},\qquad
    x_*=\fft{(3r)^{\fft13}\pi}{\sqrt{1+m^2}}.
\end{equation}
For real Chern-Simons levels $k_a$, the slope is given by $m=1/\sqrt3$.  However, if we allow $\kappa$ to be complex, $m$ can be written explicitly as
\begin{equation}
\begin{split}
	m=\begin{cases}
	\tan\fft{\pi-2\arg\kappa}{6} & (0\leq\arg\kappa\leq\pi)\\
	\tan\fft{\pi-2\arg\kappa}{6}~\text{or}~\tan\fft{-3\pi-2\arg\kappa}{6} & (-\pi<\arg\kappa<0)
	\end{cases}~.
\end{split}\label{eq:m}
\end{equation}

Turning to the subleading corrections, as we have seen in (\ref{eq:S2sub}), the first order effective action $S_2$ vanishes away from the endpoints.  We are thus left to consider the second order action $S_3$, given in (\ref{eq:Seff:3:simple}), which depends on $\rho_0(x)$, $y(x)$, $\rho_1(x)$ and $z_a(x)$.  Varying $S_3$ with respect to $\rho_1(x)$ and $z_a(x)$ gives
\begin{subequations}
\begin{align}
	0&=\fft{i}{4\pi}\fft{\kappa}{|\kappa|}(x+iy(x))^2-\fft{r\pi^2\rho_0(x)}{1+y'(x)}+\lambda_3,\label{eq:saddle:3:rho1}\\
	0&=\fft{i}{2\pi}\fft{\kappa_a}{|\kappa|}\rho_0(x)(x+iy(x))+\fft{\rho_0(x)^2(z_{a-1}(x)-2z_a(x)+z_{a+1}(x))}{1+iy'(x)}\nn\\
	&\quad-\fft{d}{dx}\fft{\pi^2\rho_0(x)^2}{2(1+iy'(x))^2},\label{eq:saddle:3:za}
\end{align}\label{eq:saddle:3:1}%
\end{subequations}
respectively, where $\lambda_3$ is a Lagrange multiplier to enforce $\rho_1$ normalization.  On the other hand, varying $S_3$ with respect to $\rho_0(x)$ and $y(x)$ gives
\begin{subequations}
\begin{align}
	0&=\fft{i}{2\pi}(x+iy(x))\sum_{a=1}^r\fft{\kappa_a}{|\kappa|}z_a(x)+\Lambda_3+\rho_0(x)\left(-\fft{\sum_{a=1}^r(z_a(x)-z_{a+1}(x))^2}{(1+iy'(x))}+\fft{\pi^2\sum_{a=1}^rz_a'(x)}{(1+iy'(x))^2}\right)\nn\\
	&\quad-\fft{r\pi^4\rho_0''(x)}{24(1+iy'(x))^3}-\fft{d^2}{dx^2}\fft{r\pi^4\rho_0(x)}{24(1+iy'(x))^3}-\fft{r\pi^2\rho_1(x)}{1+iy'(x)}+\mathcal O(y^{(n\geq2)}(x)),\label{eq:saddle:3:rho0}\\
	0&=-\fft{1}{2\pi}\fft{\kappa}{|\kappa|}\rho_1(x)(x+iy(x))-\fft{1}{2\pi}\rho_0(x)\sum_{a=1}^r\fft{\kappa_a}{|\kappa|}z_a(x)-\fft{d}{dx}\fft{i\rho_0(x)^2\sum_{a=1}^r(z_a(x)-z_{a+1}(x))^2}{2(1+iy'(x))^2}\nn\\
	&\quad+\fft{d}{dx}\fft{\pi^2i\rho_0(x)^2\sum_{a=1}^rz_a'(x)}{(1+iy'(x))^3}-\fft{d^3}{dx^3}\fft{r\pi^4i\rho_0(x)^2}{24(1+iy'(x))^4}+\fft{d^2}{dx^2}\fft{r\pi^4i\rho_0(x)\rho_0'(x)}{8(1+iy'(x))^4}\nn\\
	&\quad-\fft{d}{dx}\fft{r\pi^4i\rho_0(x)\rho_0''(x)}{8(1+iy'(x))^4}-\fft{d}{dx}\fft{r\pi^2i\rho_0(x)\rho_1(x)}{(1+iy'(x))^2}+\mathcal O(y^{(n\geq2)}(x)),\label{eq:saddle:3:y}
\end{align}\label{eq:saddle:3:2}%
\end{subequations}
respectively, where $\Lambda_3$ is a Lagrange multiplier for $\rho_0$ normalization.

To solve these equations, we first note that (\ref{eq:saddle:3:rho1}) is equivalent to the leading order saddle point equation (\ref{eq:saddle:1:rho0}) provided $\lambda_3=\Lambda_1$. Then (\ref{eq:saddle:3:za}) can be simplified by substituting in the leading order equation (\ref{eq:saddle:1:y}).  The result is
\begin{equation}
	z_{a-1}(x)-2z_a(x)+z_{a+1}(x)=-\fft{i}{2\pi}\fft{\kappa_a-\fft{\kappa}{r}}{|\kappa|}\fft{(1+iy'(x))(x+iy(x))}{\rho_0(x)},\label{eq:z-}
\end{equation}
or equivalently
\begin{equation}
\begin{split}
	z_{a-1}(x)-z_a(x)&=-\fft{i}{2\pi}\fft{r\kappa_a+(r-1)\kappa_{a+1}+\cdots+\kappa_{a+r-1}-\fft{r+1}{2}\kappa}{r|\kappa|}\\
	&\quad\times\fft{(1+iy'(x))(x+iy(x))}{\rho_0(x)}.\label{eq:z-:2}
\end{split}
\end{equation}
The remaining function we need is $\rho_1(x)$, which can be obtained from (\ref{eq:saddle:3:rho0}).  In particular, by substituting (\ref{eq:z-}) and the leading order saddle point solution (\ref{sol:saddle:1}) into (\ref{eq:saddle:3:rho0}), we find
\begin{equation}
	\rho_1(x)=\fft{1+im}{r\pi^2}\left(\Lambda_3-\fft{\pi i}{24}\fft{\kappa}{|\kappa|}\right)+\fft{1}{r(1+im)}\left(\rho_0(x)\sum_{a=1}^rz_a(x)\right)'.
\label{eq:rho1expr}
\end{equation}
The Lagrange multiplier $\Lambda_3$ can be fixed by demanding that $\rho_1(x)$ integrates to zero, as in (\ref{normalization}).  Since the integral of the total derivative in (\ref{eq:rho1expr}) vanishes due to $\rho_0(x)$ vanishing at the endpoints of the cut, we conclude that
\begin{equation}
	\Lambda_3=\fft{\pi i}{24}\fft{\kappa}{|\kappa|},
\end{equation}
so that the first term in (\ref{eq:rho1expr}) vanishes.  Finally, the last equation (\ref{eq:saddle:3:y}) is equivalent to the $x$-derivative of (\ref{eq:saddle:3:rho0}) under (\ref{sol:saddle:1}) and (\ref{eq:z-}), and hence provides no additional information.

In summary, the first sub-leading correction to the saddle point solution is given by
\begin{subequations}
\begin{empheq}[box=\fbox]{align}
    z_{a-1}(x)-z_a(x)&=-\fft{i}{2\pi}\fft{r\kappa_a+(r-1)\kappa_{a+1}+\cdots+\kappa_{a+r-1}-\fft{r+1}{2}\kappa}{r|\kappa|}\fft{(1+im)^2x}{\rho_0(x)},\\
    \rho_1(x)&=\fft{1}{r(1+im)}\left(\rho_0(x)\sum_{a=1}^rz_a(x)\right)',\label{sol:saddle:3:rho1}
\end{empheq}\label{sol:saddle:3}%
\end{subequations}
where $m$ and $\rho_0(x)$ are given in the leading order solution (\ref{sol:saddle:1}). Note that $\sum_{a=1}^rz_a(x)$ is not constrained at this order, so there is in fact a one-complex-dimensional space of solutions.  This is not expected to be a true flat direction, but will be lifted by a more careful treatment of endpoint effects.

\subsection{Subleading correction to the planar free energy}\label{sec:saddle:plannar}

Given the saddle point solution (\ref{sol:saddle:1}) and (\ref{sol:saddle:3}), we now compute the genus zero free energy by evaluating the effective action at the saddle point
\begin{equation}
    F_0=-\log Z=-N^2S_\text{cl}+o(N^2).\label{eq:F:plannar}
\end{equation}
Following the small inverse 't~Hooft parameter expansion, (\ref{eq:Seff:continuum:smallk}), we write
\begin{equation}
    S_\text{cl}=\sum_{n=1}^\infty|\kappa|^{n\alpha}S_{\text{cl},n}+o(N^0).\label{eq:Scl}
\end{equation}
where $S_{\text{cl},n}$ denotes the $|\kappa|^{n\alpha}$-order effective action $S_n(y,z_a,\rho_0,\rho_1,\ldots;\kappa_a)$ evaluated at the saddle point solution.  The leading order free energy is obtained from $S_{\mathrm{cl},1}$, while the first subleading correction comes from $S_{\mathrm{cl},3}$.  As in the saddle point determination, $S_{\mathrm{cl},2}$ vanishes at the order we are working at, although it does receive higher order endpoint corrections.

At leading order, we substitute the leading saddle point solution (\ref{sol:saddle:1}) into $S_1$ given in (\ref{eq:Seff:1:withend}) to obtain
\begin{equation}
\begin{split}
	S_{\text{cl},1}&=-\fft{3^\fft53r^\fft23\pi}{20}\fft{1-im}{\sqrt{1+m^2}}=\begin{cases}
	-\fft{3^\fft53r^\fft23\pi}{20}e^{-\fft{\pi i}{6}}(\fft{\kappa}{|\kappa|})^\fft13 & (m=\tan\fft{\pi-2\arg\kappa}{6})\\
	-\fft{3^\fft53r^\fft23\pi}{20}e^{\fft{\pi i}{2}}(\fft{\kappa}{|\kappa|})^\fft13 & (m=\tan\fft{-3\pi-2\arg\kappa}{6}).
	\end{cases}\label{eq:Seff:1:onshell}
\end{split}
\end{equation}
Here we have used the principal branch for all fractional exponents, namely 
\begin{equation}
	x^a=e^{a(\log|x|+i\arg x)}\quad (-\pi<\arg x\leq\pi).
\end{equation}
The first line in (\ref{eq:Seff:1:onshell}) applies with $m=1/\sqrt3$ for real Chern-Simons levels $k_a$.  However, there are two competing saddle point solutions for complex levels when $-\pi<\arg\kappa<0$.  This manifests itself as two options for $m$ in (\ref{eq:Seff:1:onshell}).  The actual leading order free energy is then obtained from whichever saddle has the largest real part, with the other saddle being subdominant.

In computing the leading order free energy, we ignored the contribution from the endpoint correction, $S_1^\text{end}$, which can be justified as follows. Substituting the leading order saddle point solution into $S_1^\text{end}$ given in (\ref{eq:Seff:1:end}) and rescaling $x$ appropriately, we find
\begin{equation}
\begin{split}
    S_{\text{cl},1}^\text{end}&=-\fft{r\kappa^2}{\pi^6|\kappa|^\alpha}\int_0^{|\kappa|^{-\alpha}(1+im)x_\ast}dx\,x^2(x-|\kappa|^{-\alpha}(1+im)x_\ast)^2\Big[\log\tanh x\log(1+\tanh x)\\
	&\kern10em+\text{Li}_2(1-\tanh x)+\text{Li}_2(-\tanh x)-\text{Li}_2(-1)\Big].\label{eq:Scl:1:end}
\end{split}
\end{equation}
The function within the square bracket decreases exponentially as $\sim e^{-x}$ as $\Re x$ increases. This demonstrates that the $x$-integral in (\ref{eq:Scl:1:end}) converges in the small-$|\kappa|$ limit, and moreover the leading term would be of order $\kappa^2|\kappa|^{-3\alpha}\sim\mathcal O(|\kappa|)$ at most. Hence we conclude that $S_{\text{cl},1}^\text{end}$, while only power law suppressed in small $\kappa$, is subdominant to not just $S_{\text{cl},1}$, but $S_{\text{cl},2}$ and $S_{\text{cl},3}$ as well.  Note, however, that it will be important at the next $S_{\text{cl},4}$ order, which would make its determination rather challenging.

The first subleading correction to the free energy arises from the $|\kappa|^{3\alpha}$-order classical action $S_{\text{cl},3}$. Substituting the leading, (\ref{sol:saddle:1}), and subleading, (\ref{sol:saddle:3}), saddle point solutions into the expression (\ref{eq:Seff:3:simple}) for $S_3$ and integrating over $x$ gives
\begin{equation}
	S_{\text{cl},3}=-\fft{\pi i}{48}\fft{\kappa}{|\kappa|}\left(1+\fft{12\sum_{a=1}^r(r\kappa_a+(r-1)\kappa_{a+1}+\cdots+\kappa_{a+r-1}-\fft{r+1}{2}\kappa)^2}{r\kappa^2}\right).
\end{equation}
Adding this to the leading order expression, (\ref{eq:Seff:1:onshell}), and taking into account the sign in our definition of the free energy,  (\ref{eq:F:plannar}), we find
\begin{empheq}[box=\fbox]{equation}
\begin{split}
    F_0&=N^2\bigg[\fft{3^\fft53r^\fft23\pi}{20}f(\kappa)+\fft{\pi i}{48}\kappa\Big(1+\fft{12\sum_{a=1}^r(\sum_{j=0}^{r-1}(r-j)\kappa_{a-j}-\fft{r+1}{2}\kappa)^2}{r\kappa^2}\Big)\\
    &\kern3em+\mathcal O(|\kappa|^{4\alpha})\bigg]+o(N^2),
\end{split}\label{eq:F:plannar:saddle}
\end{empheq}
where $f(\kappa)$ is given as
\begin{equation}
    f(\kappa)=\begin{cases}
    e^{-\fft{\pi i}6}\kappa^\fft13 & (\arg\kappa\in(-\pi,-\fft{\pi}{2})\cup[0,\pi])\\
    e^{\fft{\pi i}2}\kappa^\fft13 & (\arg\kappa\in(-\fft\pi2,0))\\
    \fft12|\kappa|^\fft13 & (\arg\kappa=-\fft\pi2)
    \end{cases}.
\end{equation}
Note that this expression becomes real for $\arg\kappa_a=\pm{\pi}/{2}$, which is expected from the reality of the matrix integral, (\ref{eq:Z2}), for imaginary Chern-Simons levels.

For the $r=2$ node GT theory with real Chern-Simons levels $k_1$ and $k_2$, the above expression for the free energy reduces to
\begin{equation}
    F_0=N^2\left[\fft{3^{\fft53}2^{\fft23}\pi}{20}e^{-\fft{\pi i}6}\kappa^{1/3}+\fft{\pi i}{48}\kappa\left(1+3\left(\fft{k_1-k_2}{k_1+k_2}\right)^2\right)+\cdots\right]+o(N^2),
\label{eq:GT2F0}
\end{equation}
where the inverse 't~Hooft parameter $\kappa=(k_1+k_2)/N$ is held fixed.  Here the term linear in $\kappa$ is purely imaginary and does not contribute to the real part of the free energy.  Note that, by substituting in the explicit form of $\kappa$, we can rewrite the free energy in a form suggestive of a fixed $k$ expansion
\begin{equation}
    F_0=\fft{3^{\fft53}2^{\fft23}\pi}{20}e^{-\fft{\pi i}6}N^{5/3}(k_1+k_2)^{1/3}+\fft{\pi i}{48}N(k_1+k_2)\left(1+3\left(\fft{k_1-k_2}{k_1+k_2}\right)^2\right)+\cdots.\label{eq:F:plannar:saddle:k-fix}
\end{equation}
Of course, this assumes that the expression remains valid at fixed $k$.  This will be confirmed below when expanding in the Fermi-gas approach, although additional non-planar contributions may arise that complicate the expansion at subleading order.

Here we compare our sub-leading analysis for the planar free energy of GT theory in the large 't Hooft parameter expansion with a similar analysis for that of ABJM theory \cite{Mezei:2013gqa}:
\begin{itemize}
	\item For the ABJM theory with $r=2$ and $k_1=-k_2$, since the sum of CS levels vanishes, the large 't Hooft parameter limit is taken as $N/k_1\gg1$ \cite{Mezei:2013gqa}. Here we use $N/\sum_{a=1}^rk_a\gg1$ with a non-vanishing $\sum_{a=1}^rk_a$, which can be applied to generic multi-node $(r\geq 2)$ cases in a straightforward way. 
	\item In the ABJM case, sub-leading corrections to the eigenvalue density does not affect the first sub-leading correction of the planar free energy in the large 't Hooft parameter expansion \cite{Mezei:2013gqa}. In the GT-like case, we found a similar property that the first sub-leading correction to the eigenvalue density, namely $\rho_1(x)$, corresponds to a flat direction for the planar free energy. This is a consequence of the sub-leading saddle point solution (\ref{sol:saddle:3}) that gives $\rho_1(x)$ as a total derivative, however, which is not trivial at first glance.
\end{itemize}
%

\subsection{Additional corrections beyond the planar limit}\label{sec:saddle:beyond}

Although we have focused on the genus zero free energy, it is worth considering how the saddle point expansion can be carried out beyond leading order.  For this purpose, we note the saddle point evaluation of the partition function (\ref{eq:Zr:Seff}) takes the form
\begin{equation}
    Z(N,k_a)=\fft{e^{N^2S_{\mathrm{cl}}}}{(\sqrt{2\pi}N)^{Nr}}\det\left(-\fft{\partial^2 S_{\mathrm{eff}}}{\partial\lambda_{a,i}\partial\lambda_{b,j}}\right)^{-\fft12}\left(1+\cdots\right),\label{eq:Zr:saddle}
\end{equation}
where `$\cdots$' stands for higher order corrections beyond the 1-loop determinant.  Here, the determinant arises from the Gaussian integral and the $1/(N!)^r$ prefactor in (\ref{eq:Zr:Seff}) is canceled by the degeneracy of the eigenvalues under permutations.  Taking a log then gives the free energy as
\begin{equation}
    F=-N^2 S_\text{cl}+rN\log N+\fft{rN}{2}\log2\pi+\fft12\log\det(-\fft{\partial^2 S_\text{eff}}{\partial\lambda_{a,i}\partial\lambda_{b,j}})+\mathcal O(N^0),\label{eq:F:beyond}
\end{equation}
where the planar contribution to the classical action $S_{\mathrm{cl}}$ is given in (\ref{eq:F:plannar:saddle}).

Computing the classical action and the one-loop determinant beyond the planar limit is highly involved so the details are relegated to Appendix \ref{App:saddle:beyond}, focusing on the two node $(r=2)$ case. The additional corrections include a non-planar contribution to $S_{\mathrm{cl}}$, (\ref{eq:Scl:beyond:simple}), and a determinant contribution, (\ref{eq:one-loop:beyond:simple}), with the result
\begin{empheq}[box=\fbox]{equation}
    F=F_0-r(1-\ft12\log2\pi)N+o(N),\label{eq:F:beyond:saddle}
\end{empheq}
where $F_0$ is given in (\ref{eq:F:plannar:saddle}). (While the computation in Appendix~\ref{App:saddle:beyond} focuses on $r=2$, one can see that it generalizes to an arbitrary number of nodes, $r$.)

Note that the $rN\log N$ term in (\ref{eq:F:beyond}) is canceled in the final result as expected. The origin of the linear $N$ term in (\ref{eq:F:beyond:saddle}) is still unclear. In the 't~Hooft expansion, we do not expect a linear $N$ term as sub-leading corrections start at genus one, corresponding to $\mathcal O(N^0)$ in the large-$N$ expansion.  We thus expect this term to be an artifact of the saddle point evaluation rather than a physically meaningful contribution.  It is unlikely that it would be canceled by a non-perturbative contribution, as that would normally be of $\mathcal O(\log N)$ at most. However, the cancellation may come from endpoint corrections that can mix up orders in a perturbative $1/N$ expansion.  Note that similar linear $N$ terms have been observed in the saddle point expansion of related models \cite{Liu:2018bac,Liu:2019tuk}.  We will also encounter a linear $N$ term in the Fermi-gas approach to the $S^3$ partition function for the two-node GT theory \cite{Gaiotto:2009mv} corresponding to $r=2$ in (\ref{eq:F:beyond:saddle}). As we will see below, however, it is different from the universal linear $N$ term in (\ref{eq:F:beyond:saddle}), and we comment on such a linear term separately in section \ref{sec:discussion}.

\section{The free energy from the Fermi-gas model}\label{sec:Fermi}

We now turn to the Fermi-gas approach to computing the free energy that was pioneered in \cite{Marino:2011eh}.  While the saddle point approach easily applies to $r$-node necklace quivers, the situation is less straightforward in the Fermi-gas picture.  As shown in \cite{Marino:2011eh}, the Fermi-gas approach can be readily applied to more general ABJM-like necklace quivers with matter.  However, it requires more care when considering GT-like necklace quivers. We thus restrict our analysis to the two-node GT case \cite{Gaiotto:2009mv}.

As demonstrated in \cite{Marino:2011eh}, the starting point of the Fermi-gas approach is the rewriting of the GT model partition function (\ref{eq:Z2}) as a partition function of an ideal one-dimensional Fermi-gas
\begin{empheq}[box=\fbox]{equation}
	Z(N;\theta,\Delta)=\fft{1}{N!}\sum_{\sigma\in S_N}(-1)^{\epsilon(\sigma)}\int d^Nx\prod_{i=1}^N\rho(x_i,x_{\sigma(i)};\theta,\Delta)
\label{fermi:gas:ptn:fct}
\end{empheq}
where 
\begin{equation}
    2\pi i\theta=-\fft{1}{k_1}-\fft{1}{k_2},\qquad \fft{4\pi}{\Delta}=\fft{1}{k_1}-\fft{1}{k_2}.
\label{eq:thetaDelta}
\end{equation}
The position space density matrix $\rho(x_1,x_2;\theta,\Delta)$ has the form
\begin{subequations}
	\begin{align}
	\rho(x_1,x_2;\theta,\Delta)&\equiv e^{-\fft12U(x_1;\theta)}K(x_1,x_2;\theta,\Delta)e^{-\fft12U(x_2;\theta)},\label{eq:rho:posi}\\
	U(x;\theta)&\equiv\log (2\cosh\fft{x}{2})+\fft{\theta}{2}x^2,\label{eq:U:posi}\\
	K(x_1,x_2;\theta,\Delta)&\equiv(1+\fft14\Delta^2\theta^2)^\fft12\int\fft{dy}{4\pi|\Delta|\cosh\fft{y}{2}}e^{-\fft{\theta}{2}y^2-\fft{\theta}{2}(x_1+x_2)y+\fft{i}{\Delta}(x_1-x_2)y}.\label{eq:K:posi}
	\end{align}\label{eq:posi}%
\end{subequations}
While the partition function, (\ref{fermi:gas:ptn:fct}), is given in terms of a density matrix, we can introduce the Hamiltonian
\begin{equation}
	\hat H(\theta,\Delta)\equiv-\log\hat\rho(\theta,\Delta),\label{eq:H}
\end{equation}
where the density matrix is defined in terms of its position representation
\begin{equation}
	\langle x_1|\hat\rho(\theta,\Delta)|x_2\rangle=\rho(x_1,x_2;\theta,\Delta).\label{def:rho}
\end{equation}
Note that the density matrix is \underline{not} normalized here so its trace is not unity but rather the one-particle partition function, namely $\tr\hat\rho(\theta,\Delta)=Z(1;\theta,\Delta)$.

Note that the density matrix $\hat\rho(\theta,\Delta)$ can also be written in terms of composite operators $\hat U(\theta)$ and $\hat T(\theta,\Delta)$ as
\begin{equation}
	\hat\rho(\theta,\Delta)=e^{-\hat H(\theta,\Delta)}=e^{-\fft12\hat U(\theta)}e^{-\hat T(\theta,\Delta)}e^{-\fft12\hat U(\theta)},
\label{eq:rho:composite}
\end{equation}
where $\hat T(\theta,\Delta)\equiv-\log\hat K(\theta,\Delta)$, and $\hat U(\theta)$ and $\hat K(\theta,\Delta)$ are defined in terms of their position space representations
\begin{subequations}
	\begin{align}
	\langle x_1|\hat U(\theta)|x_2\rangle&= U(x_1;\theta)\delta(x_1-x_2),\label{eq:U}\\
	\langle x_1|\hat K(\theta,\Delta)|x_2\rangle&= K(x_1,x_2;\theta,\Delta),\label{eq:K}
	\end{align}\label{eq:composite}%
\end{subequations}
respectively. Our convention for operators and their representations follows \cite{Marino:2011eh} and is summarized in Appendix \ref{App:Fermi:trace}. 

\subsection{Computing the free energy}\label{sec:Fermi:strategy}

Even with a Fermi-gas interpretation, computing the free energy $F=-\log Z$ directly from the expression (\ref{fermi:gas:ptn:fct}) for the partition function is still non-trivial.  However, Ref.~\cite{Marino:2011eh} provided a strategy for the computation, which we review here before extending the result to subleading order.

The key idea is to note that the partition function $Z(N;\theta,\Delta)$ is a \underline{canonical} partition function for fixed particle number $N$. Instead of computing this directly, it turns out to be more convenient to start with the grand canonical partition function $\Xi(\mu;\theta,\Delta)$ and the corresponding grand potential $J(\mu;\theta,\Delta)$
\begin{equation}
	e^{J(\mu;\theta,\Delta)}\equiv\Xi(\mu;\theta,\Delta)=1+\sum_{N=1}^\infty Z(N;\theta,\Delta)e^{\mu N}.
\end{equation}	
The inverse relation is then obtained by Cauchy's integral formula as
\begin{empheq}[box=\fbox]{equation}
	Z(N;\theta,\Delta)=\int_{\mu_0-\pi i}^{\mu_0+\pi i}\fft{d\mu}{2\pi i}\exp[J(\mu;\theta,\Delta)-\mu N],\label{J:to:Z}
\end{empheq}
where $\mu_0$ is an arbitrary complex number. Note that $\Xi(\mu;\theta,\Delta)$ and $J(\mu;\theta,\Delta)$ are $2\pi i$-periodic functions of $\mu$ by definition. 

Our goal now is to compute the grand potential $J(\mu;\theta,\Delta)$.  Given a spectrum for the Hamiltonian
\begin{equation}
	\hat H|\phi_n\rangle=E_n|\phi_n\rangle\qquad(n=0,1,\ldots),
\end{equation}
and taking Fermi statistics into account, the grand canonical partition function is given as 
\begin{equation}
	\Xi(\mu;\theta,\Delta)=\prod_{n\ge0}(1+e^{\mu-E_n}),
\end{equation}
or equivalently
\begin{equation}
    J(\mu;\theta,\Delta)=\sum_{n\ge0}\log(1+e^{\mu-E_n}).
\end{equation}
The discrete sum over eigenstates can be converted into an integral over continuous energies by introducing a step function $\theta(E-E_n)$ as
\begin{equation}
	J(\mu;\theta,\Delta)=\int dE\,\sum_{n\ge0}\fft{d\theta(E-E_n)}{dE}\log(1+e^{\mu-E}).\label{n:to:J:prep}
\end{equation}
This is just an integral over a density of states
\begin{empheq}[box=\fbox]{equation}
	J(\mu;\theta,\Delta)=\int dE\,\fft{dn(E;\theta,\Delta)}{dE}\log(1+e^{\mu-E}).
\label{n:to:J}
\end{empheq}
Here $n(E;\theta,\Delta)$ is the trace of the number of states operator
\begin{equation}
    n(E;\theta,\Delta)=\Tr\hat n(E;\hat H(\theta,\Delta)),
\end{equation}
where
\begin{equation}
	\hat n(E;\hat H(\theta,\Delta))\equiv\theta(E-\hat H(\theta,\Delta)).
\label{eq:n}
\end{equation}

The strategy for computing the free energy is now clear.  First, we compute the number of states $n(E;\theta,\Delta)$ below energy $E$.  This is done in the thermodynamic limit, but can be expanded to include subleading corrections. Then we compute the grand potential $J(\mu;\theta,\Delta)$ using (\ref{n:to:J}) and finally the canonical partition function $Z(N;\theta,\Delta)$ using (\ref{J:to:Z}). The free energy is then simply given as $F=-\log Z$.

\subsection{The Wigner transformed Hamiltonian}

The number of states is given by the trace of the number of states operator, $\hat n(E;\hat H(\theta,\Delta))$, given in (\ref{eq:n}).  Since we are interested in quantum corrections, we apply the semi-classical expansion of the trace of a function $f(\hat A)$ of an operator $\hat A$ as worked out in (\ref{eq:tr:f(A):semi}).  The result can be expressed as
\begin{empheq}[box=\fbox]{equation}
\begin{split}
	n(E;\theta,\Delta)&=\fft{1}{\hbar}\int\fft{dpdq}{2\pi}\theta(E-H_W(p,q;\theta,\Delta))\\
	&\quad+\hbar\int\fft{dpdq}{2\pi}\left(\fft12\delta^{(1)}(E-H_W(p,q;\theta,\Delta))\mathcal H_2^{(1)}\right.\\
	&\kern7em~\left.-\fft16\delta^{(2)}(E-H_W(p,q;\theta,\Delta))\mathcal H_3^{(1)}\right)+\mathcal O(\hbar^3),\label{eq:n(E)}
\end{split}
\end{empheq}
where $H_W(p,q;\theta,\Delta)$ is the Wigner transformed Hamiltonian and the first few $\mathcal H_n^{(j)}$ are given as 
\begin{subequations}
\begin{align}
	\mathcal H^{(1)}_2&=\fft14((\partial_q\partial_pH_W)^2-\partial_q^2H_W\partial_p^2H_W),\\
	\mathcal H^{(1)}_3&=\fft14(2\partial_qH_W\partial_pH_W\partial_q\partial_pH_W-(\partial_qH_W)^2\partial_p^2H_W-\partial_q^2H_W(\partial_pH_W)^2),
\end{align}\label{H:hbar:square}%
\end{subequations}
following (\ref{A:hbar:square}). Refer to (\ref{def:Anj}) for the definition of a generic $\mathcal H_n^{(j)}$. For notational convenience, we omit arguments of various operators and their Wigner transforms unless there are any ambiguities.

In order to proceed, we first have to compute the Wigner transform of the Hamiltonian, (\ref{eq:rho:composite}).  Since $\hat H$ is given as a product of operators, we make use of the product expression, (\ref{Wigner:product}), for the Wigner transform to write
\begin{equation}
\begin{split}
	e_\star^{-H_W}=(e^{-\hat H})_W&=(e^{-\fft12\hat U})_W\star(e^{-\hat T})_W\star(e^{-\fft12\hat U})_W\\
	&=e_\star^{-\fft12 U_W}\star e_\star^{-T_W}\star e_\star^{-\fft12 U_W},
\label{eq:rho:Wigner}
\end{split}
\end{equation}
where the $\star$ operator is defined in (\ref{star:oper}).  Applying the BCH formula (\ref{BCH}) to the RHS of (\ref{eq:rho:Wigner}), we can write down the Wigner transform of the Hamiltonian $H_W$ in terms of those of the composite operators $T_W$ and $U_W$ as 
\begin{equation}
\begin{split}
	H_W&=T_W+U_W+\fft{1}{12}[T_W,[T_W,U_W]_\star]_\star+\fft{1}{24}[U_W,[T_W,U_W]_\star]_\star+\mathcal O(\hbar^4).\label{eq:H:Wigner:general}
\end{split}
\end{equation}
If $U_W$ is a function of position $q$ only ($p$-independent), which is the case for the GT theory as we will see below in (\ref{eq:U:Wigner}), we can expand out the $\star$-commutator, (\ref{commutator}), to rewrite $H_W$ explicitly as
\begin{equation}
\begin{split}
	H_W&=T_W(p,q)+U_W(q)\\
	&\quad+\fft{\hbar^2}{24}\left(U_W'^2\partial_p^2T_W+2U_W'\partial_qT_W\partial_p^2T_W-2U_W''(\partial_pT_W)^2-2U_W'\partial_pT_W\partial_p\partial_qT_W\right)+\mathcal O(\hbar^4).\label{eq:H:Wigner}
\end{split}
\end{equation}

To compute $H_W$, we need the Wigner transforms of the composite operators, $U_W$ and $T_W$. Using the position representations (\ref{eq:posi}) and (\ref{eq:composite}) and the definition of the Wigner transformation (\ref{Wigner}), $U_W$ is easily obtained as
\begin{equation}
	U_W(q;\theta)=\log(2\cosh\fft{q}{2})+\fft{\theta}{2}q^2.\label{eq:U:Wigner}
\end{equation}
The $\star$-exponential of $T_W$ can also be obtained as
\begin{equation}
	e_\star^{-T_W(p,q;\theta,\Delta)}=K_W(p,q;\theta,\Delta)
	=\fft{(1+\fft14\Delta^2\theta^2)^\fft12}{2\cosh(\fft{\Delta}{\hbar}\fft{p}{2})}e^{-\fft{\theta}{2}(\fft{\Delta}{\hbar})^2p^2-\theta\fft{\Delta}{\hbar}pq}.\label{eq:K:Wigner}
\end{equation}
However, for (\ref{eq:H:Wigner}), we really want the $\star$-log of this expression. To do so, consider an alternative expression for $e_\star^{-T_W}$ using the Taylor expansion, namely
\begin{equation}
\begin{split}
	e_\star^{-T_W}=(e^{-\hat T})_W&=e^{-T_W}\left[\sum_{n=0}^\infty\fft{(-1)^n}{n!}(\hat T-T_W)^n\right]_W\\
	&=e^{-T_W}\left(1+\fft{1}{2!}\hbar^2\mathcal T^{(1)}_2-\fft{1}{3!}\hbar^2\mathcal T^{(1)}_3+\mathcal O(\hbar^4)\right).\label{eq:K:Wigner:2}
\end{split}
\end{equation}
Here the first few $\mathcal T^{(j)}_n$ are given as
\begin{subequations}
	\begin{align}
	\mathcal T^{(1)}_2&=\fft14((\partial_q\partial_pT_W)^2-\partial_q^2T_W\partial_p^2T_W),\\
	\mathcal T^{(1)}_3&=\fft14(2\partial_qT_W\partial_pT_W\partial_q\partial_pT_W-(\partial_qT_W)^2\partial_p^2T_W-\partial_q^2T_W(\partial_pT_W)^2),
	\end{align}\label{T:hbar:square}%
\end{subequations}
following (\ref{A:hbar:square}). Refer to (\ref{def:Anj}) for the definition of a generic $\mathcal T_n^{(j)}$. Matching (\ref{eq:K:Wigner}) with (\ref{eq:K:Wigner:2}) then gives
\begin{equation}
\begin{split}
	T_W&=\fft{\theta}{2}\left(\fft{\Delta}{\hbar}\right)^2p^2+\theta\left(\fft{\Delta}{\hbar}\right)pq+\log(2\cosh(\fft{\Delta}{\hbar}\fft{p}{2}))-\fft12\log(1+\fft14\Delta^2\theta^2)\\
	&\quad+\log(1+\fft{1}{2!}\hbar^2\mathcal T^{(1)}_2-\fft{1}{3!}\hbar^2\mathcal T^{(1)}_3+\mathcal O(\hbar^4)).\label{eq:T:Wigner:prep}
\end{split}
\end{equation}
While this expression has $T_W$ on both sides because of the $\mathcal T^{(j)}_n$, we can obtain a perturbative solution for $T_W$ in the semi-classical limit with small $\hbar$ by identifying $\Delta$ with $\hbar$ as
\begin{equation}
	\Delta=\hbar,\label{identification}
\end{equation}
assuming $\Delta>0$. The small-$\hbar$ expansion of $T_W$ is then given from (\ref{eq:T:Wigner:prep}) as
\begin{subequations}
\begin{align}
	T_W&=T_W^{(0)}+\hbar^2 T_W^{(1)}+\mathcal O(\hbar^4),\\
	T_W^{(0)}&=\fft{\theta}{2}p^2+\theta pq+\log(2\cosh\fft{p}{2}),\label{eq:T:Wigner:0}\\
	T_W^{(1)}&=-\fft{1}{8}\theta^2+\fft18\left((\partial_q\partial_pT_W^{(0)})^2-\partial_q^2T_W^{(0)}\partial_p^2T_W^{(0)}\right)\nn\\
	&\quad-\fft{1}{24}\left(2\partial_qT_W^{(0)}\partial_pT_W^{(0)}\partial_q\partial_pT_W^{(0)}-(\partial_qT_W^{(0)})^2\partial_p^2T_W^{(0)}-\partial_q^2T_W^{(0)}(\partial_pT_W^{(0)})^2\right).\label{eq:T:Wigner:1}
\end{align}\label{eq:T:Wigner}%
\end{subequations}

Finally, substituting the Wigner transforms of the composite operators $U_W$ (\ref{eq:U:Wigner}) and $T_W$ (\ref{eq:T:Wigner}) into the Wigner transformed Hamiltonian (\ref{eq:H:Wigner}) gives 
\begin{subequations}
	\begin{empheq}[box=\fbox]{align}
	H_W&=H_W^{(0)}+\hbar^2H_W^{(1)}+\mathcal O(\hbar^4),\\
	H_W^{(0)}&=\log(2\cosh\fft{p}{2})+\log(2\cosh\fft{q}{2})+\fft{\theta}{2}(p+q)^2,\\
	H_W^{(1)}&=-\fft18\theta^3(p+q)^2+\fft{1}{384}\sech^2\fft{p}{2}(2\theta(p+q)+\tanh\fft{q}{2})^2-\fft{\theta}{48}\tanh^2\fft{p}{2}\nn\\
	&\quad-\fft{1}{192}\sech^2\fft{q}{2}(2\theta(p+q)+\tanh\fft{p}{2})^2+\fft{\theta}{96}\tanh^2\fft{q}{2}\nn\\
	&\quad-\fft{\theta}{48}\tanh\fft{p}{2}(6\theta(p+q)+\tanh\fft{q}{2}).\label{eq:H:Wigner:1quant}
	\end{empheq}\label{eq:H:Wigner:semi}%
\end{subequations}
Given $H_W$, we can now compute the number of states using (\ref{eq:n(E)}). Note that a non-vanishing $\theta$, which characterizes GT-like necklace quivers, makes the Wigner transformed Hamiltonian (\ref{eq:H:Wigner:semi}) more involved and thereby the calculation of the number of states would be more complicated compared to the ABJM-like case.

\subsection{Computing the number of states}
\label{sec:Fermi:calculation:number}
Since the Wigner transformed Hamiltonian $H_W$ is obtained perturbatively with respect to $\hbar^2$, the number of states $n(E;\theta,\hbar)$ in (\ref{eq:n(E)}) can also be written perturbatively as
\begin{empheq}[box=\fbox]{equation}
	n(E;\theta,\hbar)=\fft{1}{\hbar}\left(n^{(0)}(E;\theta)+\hbar^2 n^{(1)}(E;\theta)+\mathcal O(\hbar^4)\right),\label{eq:n(E):semi}
\end{empheq}
where the classical contribution is given as
\begin{empheq}[box=\fbox]{equation}
	n^{(0)}(E;\theta)=\int\fft{dpdq}{2\pi}\,\theta(E-H_W^{(0)}(p,q;\theta)),\label{eq:n(E):classic}
\end{empheq}
and the first quantum correction is given as
\begin{empheq}[box=\fbox]{equation}
\begin{split}
	&n^{(1)}(E;\theta)\\
	&=\underbrace{\int\fft{dpdq}{2\pi}\,\delta(E-H_W^{(0)})\big(-H_W^{(1)}\big)}_{n_1^{(1)}(E;\theta)}\\
	&\quad+\underbrace{\int\fft{dpdq}{2\pi}\left(\fft12\delta^{(1)}(E-H_W^{(0)})\mathcal H_2^{(1)}-\fft16\delta^{(2)}(E-H_W^{(0)})\mathcal H_3^{(1)}\right)}_{n_2^{(1)}(E;\theta)}.\label{eq:n(E):quant}
\end{split}
\end{empheq}
Here $\mathcal H_n^{(j)}$ is given by (\ref{H:hbar:square}), but with $H_W$ replaced by $H_W^{(0)}$.

\subsubsection{The classical contribution}

We begin with the classical contribution $n^{(0)}(E;\theta)$ to the number of states.  As in (\ref{eq:n(E):classic}), this is just the area of the region bounded by the Fermi surface $E=H_W^{(0)}(p,q;\theta)$ in phase space divided by $2\pi$.  Note that, from here on, we assume \underline{real positive} $\theta$. One immediate consequence is that $n^{(0)}(E;\theta)=0$ for $E\leq 2\log2$ since
\begin{equation}
	H_W^{(0)}(p,q;\theta)=\log(2\cosh\fft{p}{2})+\log(2\cosh\fft{q}{2})+\fft{\theta}{2}(p+q)^2\geq2\log2\label{lower:bdd}
\end{equation}
for any real $p,q$. 

To compute the area, we first make the change of variables
\begin{equation}
	u=\fft{1}{\sqrt2}(p+q),\qquad v=\fft{1}{\sqrt2}(p-q),
\end{equation}
so that
\begin{equation}
    H_W^{(0)}(u,v;\theta)=\theta u^2+\log(2\cosh\fft{u}{\sqrt2}+2\cosh\fft{v}{\sqrt2}).
\label{eq:Hv}
\end{equation}
Then (\ref{eq:n(E):classic}) can be rewritten as
\begin{equation}
	n^{(0)}(E;\theta)=\fft{1}{2\pi}\oint_{H_W^{(0)}(u,v;\theta)=E}v\,du=\fft{2}{\pi}\int_0^{u_\ast}v^{(0)}(u;E,\theta)\,du,
\label{eq:n(E):classic:1}
\end{equation}
where
\begin{equation}
	v^{(0)}(u;E,\theta)=\sqrt2\cosh^{-1}\left[\fft12e^{E-\theta u^2}-\cosh\fft{u}{\sqrt2}\right],
\label{eq:v}
\end{equation}
is obtained by inverting $H_W^{(0)}(u,v;\theta)=E$.  The end point $u_\ast$ is determined by the constraint $v^{(0)}(u_\ast;E,\theta)=0$, and by symmetry we have integrated over one quadrant and multiplied the result by four.

The difficulty with computing the integral (\ref{eq:n(E):classic:1}) lies in the expression for $v^{(0)}(u;E,\theta)$.  To proceed, we work in the thermodynamic limit $E\gg1$.  In this case, the $e^{E-\theta u^2}$ term dominates in (\ref{eq:v}), so we pull it out and write
\begin{equation}
	n^{(0)}(E;\theta)=\underbrace{\fft{2\sqrt2}{\pi}\int_0^{u_\ast}du\,(E-\theta u^2)}_{n_1^{(0)}(E;\theta)}+\underbrace{\fft{2\sqrt2}{\pi}\int_0^{u_\ast}du\log(1-g(u;\theta))}_{n_2^{(0)}(E;\theta)},\label{eq:n(E):classic:2}
\end{equation}
where we have defined the function $g(u)$ as
\begin{align}
	f(u;\theta)&\equiv e^{-(E-\theta u^2)}\cosh\fft{u}{\sqrt2},\label{eq:f}\\
	g(u;\theta)&\equiv \fft12+f(u;\theta)-\sqrt{\left(\fft12-f(u;\theta)\right)^2-e^{-2(E-\theta u^2)}}.\label{eq:g}
\end{align}
The first term, $n^{(0)}_1(E;\theta)$, in the large-$E$ expansion was considered in \cite{Marino:2011eh} where $u_*$ was further approximated by the large-$E$ result $u_*\approx\sqrt{E/\theta}$.  However, since we are interested in subleading corrections, we need a more complete expression for $u_*$.  In particular, for $E>{1}/{8\theta}$, we can expand $u_\ast$ in a series up to exponentially suppressed terms
\begin{equation}
	u_\ast=\fft{-1+\sqrt{1+8\theta E}}{2\sqrt2\theta}-\sqrt{\fft{8}{1+8\theta E}}e^{\fft{1-\sqrt{1+8\theta E}}{4\theta}}(1+\mathcal O(e^{-\sqrt{E/2\theta}})).\label{eq:u*}
\end{equation}
It is then straightforward to compute the large-$E$ expansion of $n^{(0)}_1(E;\theta)$ as
\begin{equation}
\begin{split}
	n^{(0)}_1(E;\theta)&=\fft{1+(4\theta E-1)\sqrt{1+8\theta E}}{6\pi\theta^2}-\fft{2}{\pi\theta}e^{\fft{1-\sqrt{1+8\theta E}}{4\theta}}(1+\mathcal O(E^{-\fft12}))\\
	&=\fft{4\sqrt2}{3\pi\theta^\fft12}E^\fft32-\fft{\sqrt2}{4\pi\theta^\fft32}E^\fft12+\fft{1}{6\pi\theta^2}-\fft{3\sqrt2}{128\pi\theta^\fft52}E^{-\fft12}\\
	&\quad+\sum_{n=1}^\infty c_nE^{-n-\fft12}-\fft{2}{\pi\theta}e^{\fft{1-\sqrt{1+8\theta E}}{4\theta}}(1+\mathcal O(E^{-\fft12})),\label{eq:n1(E):classic}
\end{split}
\end{equation}
where we have given the explicit expansion of the power law terms up to $\mathcal O(E^{-1/2})$.

Here we emphasize that the large-$E$ expansion of the endpoint (\ref{eq:u*}) is not allowed in the limit $\theta\to0$, which corresponds to the ABJM case with $k_1+k_2\to0$. Hence, even though what we have discussed up to the expansion (\ref{eq:u*}) reproduces the known ABJM results smoothly under the limit $\theta\to0$, once we compute the large-$E$ expansion of the number of states using the expression (\ref{eq:u*}) and derive a partition function based on it, the results will not reproduce the known ABJM results under $\theta\to0$ anymore.

We now turn to the second term in (\ref{eq:n(E):classic:2}).  To compute $n^{(0)}_2(E;\theta)$, first note that $f(u;\theta)$ in (\ref{eq:f}) is an increasing function of $u$ within the domain $u\in[0,u_\ast]$ such that $0<f(u;\theta)<\fft12$. Then we can show that $g(u;\theta)$ in (\ref{eq:g}) is also an increasing function of $u$ within the domain $u\in[0,u_\ast]$ such that $0<g(u;\theta)<1$. Hence $n^{(0)}_2(E;\theta)$ in (\ref{eq:n(E):classic:2}) can be rewritten using the Taylor expansion of $\log(1-g(u))$ as
\begin{equation}
	n^{(0)}_2(E;\theta)=-\fft{2\sqrt2}{\pi}\sum_{l=1}^\infty\fft{1}{l}\int_0^{u_\ast}du\,g(u)^l.\label{eq:n2(E):classic:1}
\end{equation}
We now further expand $g(u;\theta)$ as
\begin{equation}
\begin{split}
	g(u;\theta)&=\fft12+f(u)-\left(\fft12-f(u)\right)\sum_{n=0}^\infty\binom{1/2}{n}(-1)^nh(u)^{2n}\\
	&=e^{-(E-\theta u^2)}\left(2\cosh\fft{u}{\sqrt2}-\sum_{n=1}^\infty\binom{1/2}{n}(-1)^nh(u)^{2n-1}\right),
\end{split}\label{eq:g:Taylor}
\end{equation}
where we have defined $h(u)$ as
\begin{equation}
	h(u)\equiv\fft{e^{-(E-\theta u^2)}}{1/2-f(u)}=2e^{-(E-\theta u^2)}\sum_{m=0}^\infty\binom{-1}{m}(-2f(u))^m.
\end{equation}
Note that $h(u)$ is an increasing function within the domain $u\in[0,u_*]$ such that $0<h(u)\leq e^{-\theta((u_*)^2-u^2)}$. Substituting (\ref{eq:g:Taylor}) into (\ref{eq:n2(E):classic:1}) then gives
\begin{equation}
\begin{split}
	&n_2^{(0)}(E;\theta)\\
	&=-\fft{2\sqrt2}{\pi}\sum_{l=1}^\infty\fft{1}{l}\int_0^{u_\ast}du\,e^{-l(E-\theta u^2)+\fft{l}{\sqrt2}u}\left(1+e^{-\sqrt2 u}-e^{-\fft{u}{\sqrt2}}\sum_{n=1}^\infty\binom{1/2}{n}(-1)^nh(u)^{2n-1}\right)^l.
\end{split}
\end{equation}
Applying the integral (\ref{building:block:0}) to the above result, we obtain the large-$E$ expansion of $n_2^{(0)}(E;\theta)$ as
\begin{equation}
\begin{split}
	n_2^{(0)}(E;\theta)&=-\fft{2\sqrt2}{\pi}\left(\fft{\zeta(2)}{2}(\theta E)^{-\fft12}+\left(\fft{\theta\zeta(3)}{4}-\fft{\zeta(2)}{32}\right)(\theta E)^{-\fft32}+\mathcal O(E^{-\fft52})\right)\\
	&\quad+\mathcal O(e^{-\sqrt{E/2\theta}}).\label{eq:n2(E):classic:2}
\end{split}
\end{equation}
Here we see explicitly that $n_2^{(0)}(E;\theta)$ is small in the large-$E$ limit, and hence only contributes to subleading corrections to the free energy.

Finally, substituting (\ref{eq:n1(E):classic}) and (\ref{eq:n2(E):classic:2}) into (\ref{eq:n(E):classic:2}) gives the classical contribution to the number of states
\begin{empheq}[box=\fbox]{equation}
\begin{split}
	n^{(0)}(E;\theta)&=\fft{4\sqrt2}{3\pi\theta^\fft12}E^\fft32-\fft{\sqrt2}{4\pi\theta^\fft32}E^\fft12+\fft{1}{6\pi\theta^2}-\fft{3\sqrt2}{128\pi\theta^\fft52}\left(1+\fft{128\theta^2}{3}\zeta(2)\right)E^{-\fft12}\\
	&\quad+\sum_{n=1}^\infty c_nE^{-n-\fft12}+\mathcal O(e^{-\sqrt{E/2\theta}}),\label{eq:n(E):classic:3}
\end{split}
\end{empheq}
where $c_n$ in (\ref{eq:n1(E):classic}) is redefined to incorporate the contribution from $n_2^{(0)}(E;\theta)$ in (\ref{eq:n2(E):classic:2}).

\subsubsection{The first quantum correction}

Recall that the number of states, $n(E;\theta,\hbar)$, in (\ref{eq:n(E):semi}) receives both classical and quantum contributions.  We now consider the first quantum correction, $n^{(1)}(E;\theta)$, in (\ref{eq:n(E):quant}). To begin with, we substitute 
\begin{equation}
	\delta(E-H_W^{0}(u,v;\theta))=\fft{\delta(v-v^{(0)}(u;E,\theta))}{|\partial_vH_W^{(0)}|}+\fft{\delta(v+v^{(0)}(u;E,\theta))}{|\partial_vH_W^{(0)}|},\label{Dirac:identity}
\end{equation}
into $n_2^{(1)}(E;\theta)$ in (\ref{eq:n(E):quant}). Integrating over $v$ with the Dirac-delta function and substituting explicit values of (\ref{eq:Hv}) then gives
\begin{equation}
\begin{split}
	&n_2^{(1)}(E;\theta)\\
	&=-\fft{d}{dE}\int_0^{u_\ast}du\,\fft{2\theta\cosh\fft{u}{\sqrt2}+e^{-(E-\theta u^2)}(1-4\theta\sinh^2\fft{u}{\sqrt2})}{4\pi\sqrt2S_v^0}\\
	&\quad+\fft{d^2}{dE^2}\int_0^{u_\ast}du\,\fft{\sqrt2}{24\pi S_v^0}\left(\theta e^{E-\theta u^2}+\left(\fft{1}{2}(1-8\theta+8\theta^2u^2)\cosh\fft{u}{\sqrt2}+2\sqrt2\theta u\sinh\fft{u}{\sqrt2}\right)\right.\\
	&\kern11em+e^{-(E-\theta u^2)}\left(-1-4\theta+8\theta^2 u^2-4\sqrt2\theta u\cosh\fft{u}{\sqrt2}\sinh\fft{u}{\sqrt2}\right.\\
	&\kern17em\left.\left.+(-1+4\theta-8\theta^2 u^2)\cosh^2\fft{u}{\sqrt2}\right)\right)\\
	&=\mathcal O(E^{-\fft32}),\label{eq:n1(E):quant}
\end{split}
\end{equation}
where $S_v^0\equiv\sinh(\fft1{\sqrt2}{v^{(0)}(u;E,\theta)})$.  The last estimate in (\ref{eq:n1(E):quant}) is from various integrals in Appendix \ref{App:Fermi:integrals:2}. This expansion can be computed more explicitly, but it will not be necessary since we are working only to $\mathcal O(E^{-1/2})$.

Next we consider $n_1^{(1)}(E;\theta)$ in (\ref{eq:n(E):quant}). Substituting the identity (\ref{Dirac:identity}) and integrating over $v$ with the Dirac-delta function, we obtain
\begin{equation}
	n_1^{(1)}(E;\theta)=\fft{1}{\pi}\int_{-u_*}^{u_*} du\,(\partial_vH_W^{(0)})^{-1}(-H_W^{(1)})\big|_{v=v^{(0)}(u;E,\theta)},\label{eq:n2(E):quant:uv}
\end{equation}
where the integrand is given as
\begin{equation}
\begin{split}
	&-(\partial_vH_W^{(0)})^{-1}H_W^{(1)}\big|_{v=v^{(0)}(u;E,\theta)}\\&=\fft{\sqrt2}{8S_v^0}\theta e^{E-\theta u^2}\left(-\fft{1}{24}+\theta^2u^2\right)+\fft14\theta^2 u\\&\quad+\fft{\sqrt2}{384S_v^0}\left((1+4\theta+4\sqrt2\theta(1+6\theta)u+8\theta^2u^2)e^{\fft{u}{\sqrt2}}+(1+4\theta-4\sqrt2\theta(1+6\theta)u+8\theta^2u^2)e^{-\fft{u}{\sqrt2}}\right)\\
	&\quad-\fft{\sqrt2}{384S_v^0}e^{-(E-\theta u^2)}\left((1-4\theta+4\sqrt2\theta u+8\theta^2u^2)e^{\sqrt2 u}+2(3+4\theta-8\theta^2u^2)\right.\\
	&\kern10em\left.+(1-4\theta-4\sqrt2\theta u+8\theta^2u^2)e^{-\sqrt2 u}\right)\\
	&\quad+\fft{\sqrt2}{64}e^{-(E-\theta u^2)}\left((1+4\theta+4\sqrt2\theta u+8\theta^2u^2)e^{\fft{u}{\sqrt2}}-(1+4\theta-4\sqrt2\theta u+8\theta^2u^2)e^{-\fft{u}{\sqrt2}}\right).
\end{split}
\end{equation}
Here we have used explicit values of (\ref{eq:H:Wigner:1quant}) and (\ref{eq:Hv}). Applying the results of Appendix \ref{App:Fermi:integrals:2}, we can compute the integral (\ref{eq:n2(E):quant:uv}) explicitly as
\begin{equation}
\begin{split}
	&n_1^{(1)}(E;\theta)\\
	&=\fft{2}{\pi}\int_0^{u_\ast}du\,\left[\fft{\sqrt2}{8S_v^0}\theta e^{E-\theta u^2}\left(-\fft{1}{24}+\theta^2u^2\right)\right.\\
	&\kern6em~~+\fft{\sqrt2}{384S_v^0}(1+4\theta+4\sqrt2\theta(1+6\theta)u+8\theta^2u^2)e^{\fft{u}{\sqrt2}}\\
	&\kern6em~~\left.-\fft{\sqrt2}{384S_v^0}e^{-(E-\theta u^2)}(1-4\theta+4\sqrt2\theta u+8\theta^2u^2)e^{\sqrt2 u}\right]+\mathcal O(e^{-\sqrt E})\\
	&=\fft{\theta^\fft32}{3\sqrt2\pi}E^\fft32+\fft{5\theta^\fft12}{48\sqrt2\pi}E^\fft12+\fft{7-128\theta-384\theta^2\zeta(2)}{1536\sqrt2\pi\theta^\fft12}E^{-\fft12}+\mathcal O(E^{-\fft32}).\label{eq:n2(E):quant}
\end{split}
\end{equation}
Substituting the two contributions (\ref{eq:n1(E):quant}) and (\ref{eq:n2(E):quant}) into (\ref{eq:n(E):quant}) then gives the first quantum correction to the number of states
\begin{empheq}[box=\fbox]{equation}
	n^{(1)}(E;\theta)=\fft{\theta^\fft32}{3\sqrt2\pi}E^\fft32+\fft{5\theta^\fft12}{48\sqrt2\pi}E^\fft12+\fft{7-128\theta-384\theta^2\zeta(2)}{1536\sqrt2\pi\theta^\fft12}E^{-\fft12}+\mathcal O(E^{-\fft32}).\label{eq:n(E):quant:1}
\end{empheq}
The $E^{-\fft12}$ order term with the factor of $-128\theta$ is somewhat unexpected, and leads to an irregular term in the final result for the free energy (\ref{eq:F:Fermi}): the term containing $\hbar^2\theta^2(-16\theta)$. We will comment on this contribution further in section \ref{sec:discussion}.

In principle, we can continue to work out the higher order quantum corrections by further expanding the Wigner transformed Hamiltonian as well as the phase space integral, (\ref{eq:n(E)}), for the number of states.  However, the expressions become rather unwieldy after the first quantum corrections.  To get a general idea of the higher quantum corrections, we have constructed the $\hbar^4$-order $H_W^{(2)}$ in Mathematica and used numerical integration to determine the second quantum correction to the number of states, $n^{(2)}(E;\theta)$.  The result is
\begin{align}
    n^{(2)}(E;\theta)&=-4.68914746(1)\times10^{-3}\,\theta^{\fft72}E^{\fft32}-3.80(1)\times10^{-3}\,\theta^{\fft52}E^{\fft12}+o(E^{\fft12})\nn\\
    &\approx\fft{4\sqrt2}{3\pi\theta^{\fft12}}\left(-\fft{\theta^4}{128}\right)E^{\fft32}-\fft{\sqrt2}{4\pi\theta^{\fft32}}\left(\fft{13\theta^4}{384}\right)E^{\fft12}+o(E^{\fft12}).
\label{eq:quant2}
\end{align}
The rational coefficients in the second line are consistent with the error estimates, although the numerical errors are too large to determine the second coefficient with any level of confidence.  The reason we suggest the value $13/384$ will be justified \textit{a postiori} when we connect the Fermi-gas result to the saddle point expression, (\ref{eq:GT2F0}).

Finally, adding the classical contribution, (\ref{eq:n(E):classic:3}), the first quantum correction, (\ref{eq:n(E):quant:1}), and the numerically determined second quantum correction, (\ref{eq:quant2}), gives the large-$E$ expansion of the number of states in the semi-classical limit
\begin{empheq}[box=\fbox]{equation}
    n(E;\theta,\hbar)=\alpha E^\fft32+\beta E^\fft12+\gamma+\delta_0E^{-\fft12}+\sum_{n=1}^\infty\delta_nE^{-(n+\fft12)}+\mathcal O(e^{-\sqrt E}),
\label{eq:n(E):largeE}
\end{empheq}
where
\begin{subequations}
\begin{empheq}[box=\fbox]{align}
	\alpha&=\fft{4\sqrt2}{3\pi\hbar\theta^\fft12}\left(1+\fft{\hbar^2\theta^2}{8}-\fft{\hbar^4\theta^4}{128}+\mathcal O(\hbar^6)\right),\label{eq:alpha}\\
	\beta&=-\fft{\sqrt2}{4\pi\hbar\theta^\fft32}\left(1-\fft{5\hbar^2\theta^2}{24}+\fft{13\hbar^4\theta^4}{384}+\mathcal O(\hbar^6)\right)+\mathcal O(\hbar^3),\\
	\gamma&=\fft{1}{6\pi\hbar\theta^2}\biggl(1+\mathcal O(\hbar^4)\biggr),\\
	\delta_0&=-\fft{3\sqrt2}{128\pi\hbar\theta^\fft52}\left(\left(1+\fft{128\theta^2}{3}\zeta(2)\right)-\fft{\hbar^2\theta^2(7-128\theta-384\theta^2\zeta(2))}{72}+\mathcal O(\hbar^4)\right).
\end{empheq}\label{coefficients}%
\end{subequations}
Here $\delta_n$ ($n\ge1$) is the first quantum corrected version of $c_n$ in (\ref{eq:n(E):classic:3}) and can be computed explicitly if necessary. The $\mathcal O(\hbar^4)$ terms in $\alpha$ and $\beta$ are numerically derived.

It is worth mentioning that the large-$E$ expansion of the number of states yields an infinite series (\ref{eq:n(E):largeE}). This is distinguished from the ABJM result in terms of a finite polynomial up to non-perturbative corrections, namely \cite{Marino:2011eh}
\begin{equation}
    n_\text{ABJM}(E)=CE^2+n_0+\mathcal O(Ee^{-E}),
\end{equation}
which plays a crucial role in deriving the exact Airy function form of the free energy. We therefore expect there would be no simple closed expression for the free energy of the GT theory. However, we can still investigate the first few leading terms in the double scaling limit spelled out in (\ref{expansion:Fermi}), which will be implemented in the next two subsections.

\subsection{Computing the grand potential}\label{sec:Fermi:calculation:grand}

The large-$E$ expansion, (\ref{eq:n(E):largeE}), for the number of states $n(E;\theta,\hbar)$ is only valid for sufficiently large $E$, namely for $E\geq E_*(\theta,\hbar)$ with some bound $E_*(\theta,\hbar)$.  To be precise, we choose $E_\ast(\theta,\hbar)$ large enough so that the series $\sum_{n=1}^\infty|\delta_n|(E_\ast(\theta,\hbar))^{-(n+\fft12)}$ is absolutely convergent; this will be used in Appendix~\ref{App:Fermi:grand:potential} to bound the large chemical potential expansion.  The grand potential is now computed through the integral (\ref{n:to:J}).  However, we must integrate over all energies, so the large-$E$ expansion, (\ref{eq:n(E):largeE}), is insufficient by itself.  Since we do not have an explicit result for the number of states away from the large-$E$ limit, we have to work a bit to bound any possible corrections that arise at finite $E$.

As noted above in (\ref{lower:bdd}), the number of states vanishes in the small-$E$ regime
\begin{equation}
    n(E;\theta,\hbar)=0\quad(E<2\log 2).\label{eq:n(E):smallE}
\end{equation}
Thus, for a generic $E$ we have
\begin{equation}
\begin{split}
	n(E;\theta,\hbar)=\begin{cases}
	\alpha E^\fft32+\beta E^\fft12+\gamma+\delta_0E^{-\fft12}+\tilde n(E;\theta,\hbar) & (E\geq E_0)\\
	0 & (E<E_0)
	\end{cases},\label{eq:n(E):2}
\end{split}
\end{equation}
where $E_0=2\log2$ and $\tilde n(E;\theta,\hbar)$ is supposed to satisfy
\begin{equation}
\begin{split}
	&n(E_0;\theta,\hbar)=\alpha E_0^\fft32+\beta E_0^\fft12+\gamma+\delta_0E_0^{-\fft12}+\tilde n(E_0;\theta,\hbar)=0,\\
	&\left|\tilde n(E;\theta,\hbar)-\sum_{n=1}^\infty\delta_nE^{-(n+\fft12)}\right|\leq\mathcal A(\theta,\hbar)e^{-\mathcal B(\theta,\hbar)\sqrt{E}}\qquad(E\geq E_\ast(\theta,\hbar)),\label{ntilde:property}
\end{split}
\end{equation}
with some $E$-independent positive constants $\mathcal A(\theta,\hbar)$ and $\mathcal B(\theta,\hbar)$. Now, substituting the expansion (\ref{eq:n(E):2}) into the integral (\ref{n:to:J}) for the grand potential and using the Fermi-Dirac integral representation of the polylogarithm, namely ($\Re(s)>-1$ and $z\in\mathbb C\setminus(-\infty,-1]$)
\begin{equation}
\begin{split}
	&\int_{E_0}^\infty dE\, s\,E^{s-1}\log(1+e^{\mu-E})\\
	&=-\Gamma(s+1)\mathrm{Li}_{s+1}(-e^\mu)-\int_0^{E_0}dE\,\fft{E^s}{1+e^{E-\mu}}-E_0^s\log(1+e^{\mu-E_0}),
\end{split}
\end{equation}
we have
\begin{equation}
\begin{split}
	J(\mu;\theta,\hbar)&=-\alpha\Gamma(5/2)\mathrm{Li}_{\fft52}(-e^\mu)-\beta\Gamma(3/2)\mathrm{Li}_{\fft32}(-e^\mu)-\delta_0\Gamma(1/2)\mathrm{Li}_{\fft12}(-e^\mu)\\
	&\quad+\gamma(\mu-E_0)-\int_0^{E_0}dE\,\fft{\alpha E^\fft32+\beta E^\fft12+\delta_0 E^{-\fft12}}{1+e^{E-\mu}}+\int_{E_0}^\infty dE\,\tilde n(E;\theta,\hbar)\\
	&\quad-\int_{E_0}^\infty dE\,\fft{\tilde n(E;\theta,\hbar)}{1+e^{\mu-E}}+\mathcal O(e^{-\mu}).\label{eq:J}
\end{split}
\end{equation}
To derive (\ref{eq:J}) we have also used the properties of $\tilde n(E;\theta,\hbar)$ given in (\ref{ntilde:property}). 

We now obtain the large $\Re\mu$ expansion of the grand potential $J(\mu;\theta,\hbar)$. For the first line of (\ref{eq:J}), we use the large $\Re\mu$ expansion of the polylogarithm (see (11.1) in \cite{Wood:1992} for example),
\begin{equation}
	\mathrm{Li}_s(-e^\mu)=-2\sum_{k=0}^\infty(1-2^{1-2k})\zeta(2k)\fft{\mu^{s-2k}}{\Gamma(s+1-2k)}+\mathcal O(e^{-\mu})\quad(\Im\mu\in(-\pi,\pi]).\label{polylog:exp}
\end{equation}
In this expansion, the imaginary part of $\mu$ must be within $(-\pi,\pi]$ if we use the principal branch for $\mu^{s-2k}$.  For the integrals in the remaining lines of (\ref{eq:J}), we show in Appendix~\ref{App:Fermi:grand:potential} that they are of order $\mathcal O(\mu^0)$.  As a result, we obtain the large $\Re\mu$ expansion of (\ref{eq:J}) as
\begin{empheq}[box=\fbox]{equation}
	J(\mu;\theta,\hbar)=\fft25\alpha\mu^\fft52+\fft23\beta\mu^\fft32+\gamma\mu+\epsilon\mu^\fft12+\mathcal O(\mu^0)\qquad(\Im\mu\in(-\pi,\pi]).\label{eq:J:1}
\end{empheq}
Here we have defined $\epsilon$ as
\begin{equation}
    \epsilon\equiv\fft32\zeta(2)\alpha+2\delta_0=-\fft{3\sqrt2}{64\pi\hbar\theta^\fft52}\left(1-\fft{\hbar^2\theta^2(7-128\theta)}{72}\right)+\mathcal O(\hbar^3),\label{coefficient:e}
\end{equation}
and the remaining coefficients are given in (\ref{coefficients}).

\subsection{Computing the partition function}\label{sec:Fermi:calculation:partfcn}
Finally we are ready to compute the partition function $Z$ and therefore the corresponding free energy $F=-\log Z$. Substituting the large $\Re\mu$ expansion (\ref{eq:J:1}) into (\ref{J:to:Z}), we obtain
\begin{equation}
    Z(N;\theta,\hbar)=N^\fft23\int_{N^{-\fft23}(\mu_0-\pi i)}^{N^{-\fft23}(\mu_0+\pi i)}\fft{d\mu}{2\pi i}\exp[N^\fft53G(\mu;\theta,\hbar)],\label{eq:Z}
\end{equation}
where we have rescaled $\mu\to N^\fft23\mu$ and defined $G(\mu;\theta,\hbar)$ as
\begin{equation}
    G(\mu;\theta,\hbar)\equiv\fft25\alpha\mu^\fft52+\fft23\beta N^{-\fft23}\mu^\fft32-(1-\gamma N^{-1})\mu+\epsilon N^{-\fft43}\mu^\fft12+O(N^{-\fft53}).
\end{equation}
Recall that $\mu_0$ in (\ref{J:to:Z}) was an arbitrary complex number. Here we have chosen a large enough \underline{real} $\mu_0$, which makes the large $\Re\mu$ expansion (\ref{eq:J:1}) valid over the integration range of $\mu$. 

We compute the integral (\ref{eq:Z}) using a saddle point approximation in the large-$N$ limit. To begin with, a saddle point $\mu=\mu_\ast$ for the effective action $G(\mu;\theta,\hbar)$ is obtained by demanding $G'(\mu_\ast;\theta,\hbar)=\mathcal O(N^{-\fft53})$ as
\begin{equation}
    \mu_\ast=\alpha^{-\fft23}-\fft23\alpha^{-1}\beta N^{-\fft23}-\fft23\alpha^{-\fft23}\gamma N^{-1}+\fft19\alpha^{-\fft43}(\beta^2-3\alpha\epsilon)N^{-\fft43}+O(N^{-\fft53}).\label{eq:mu*}
\end{equation}
The saddle point evaluation of the partition function (\ref{eq:Z}) is then given as
\begin{equation}
\begin{split}
    &Z(N;\theta,\hbar)\\
    &=N^\fft23\mathrm{exp}\left[N^\fft53G(\mu_\ast)\right]\\
    &\quad\times\int_{N^{-\fft23}(\mu_0-\pi i)}^{N^{-\fft23}(\mu_0+\pi i)}\fft{d\mu}{2\pi i}\mathrm{exp}\left[N^\fft53\left(\fft{G''(\mu_*)}{2!}(\mu-\mu_\ast)^2+\sum_{n=3}^\infty\fft{G^{(n)}(\mu_\ast)}{n!}(\mu-\mu_\ast)^n\right)\right],\label{eq:Z:0}
\end{split}
\end{equation}
where the first two non-trivial Taylor expansion coefficients are given as
\begin{subequations}
	\begin{align}
	G(\mu_\ast)&=-\fft35\alpha^{-\fft23}+\fft23\alpha^{-1}\beta N^{-\fft23}+\alpha^{-\fft23}\gamma N^{-1}-\fft13\alpha^{-\fft43}(\beta^2-3\alpha\epsilon)N^{-\fft43}\nn\\
	&\qquad+O(N^{-\fft53}),\\
	G''(\mu_\ast)&=\fft32\alpha^{\fft23}-\fft12\alpha^{\fft23}\gamma N^{-1}+\fft16(\beta^2-3\alpha\epsilon)N^{-\fft43}+O(N^{-\fft53}).
	\end{align}
\end{subequations}
Now we set $\mu_0=N^\fft23\mu_\ast$, which satisfies the assumption that $\mu_0$ should be a large enough real parameter. (Recall that we have assumed $\hbar>0$ in (\ref{identification}) and therefore $\mu_\ast>0$ according to (\ref{coefficients}) and (\ref{eq:mu*}) in the large-$N$ limit.) Rescaling $\mu$ as $\mu\to\mu_\ast+N^{-\fft56}\mu$ in (\ref{eq:Z:0}) then gives
\begin{equation}
\begin{split}
	&Z(N;\theta,\hbar)\\
	&=N^{-\fft16}\mathrm{exp}\left[N^\fft53G(\mu_\ast)\right]\int_{-N^\fft16\pi i}^{N^\fft16\pi i}\fft{d\mu}{2\pi i}\mathrm{exp}\left[\fft{G''(\mu_*)}{2!}\mu^2+\sum_{n=3}^\infty\fft{G^{(n)}(\mu_\ast)}{n!}N^{\fft{10-5n}{6}}\mu^n\right].\label{eq:Z:1}
\end{split}
\end{equation}
Under the large-$N$ limit, the integral reduces to a finite number as
\begin{equation}
\begin{split}
	\lim_{N\to\infty}\int_{-N^\fft16\pi i}^{N^\fft16\pi i}\fft{d\mu}{2\pi i}\mathrm{exp}\left[\fft{G''(\mu_*)}{2!}\mu^2+\sum_{n=3}^\infty\fft{G^{(n)}(\mu_\ast)}{n!}N^{\fft{10-5n}{6}}\mu^n\right]=\fft{1}{2\pi}\sqrt{\fft{4\pi}{3\alpha^\fft23}}.
\end{split}
\end{equation}
Hence the large-$N$ expansion of the free energy $F=-\log Z$ is determined by the prefactor of (\ref{eq:Z:1}) as
\begin{equation}
	F=\fft35\alpha^{-\fft23}N^\fft53-\fft23\alpha^{-1}\beta N-\alpha^{-\fft23}\gamma N^\fft23+\fft13\alpha^{-\fft43}(\beta^2-3\alpha\epsilon)N^\fft13+\fft16\log N+O(N^{0}).
\end{equation}
Substituting actual values of $\alpha,\beta,\gamma,\epsilon$ given in (\ref{coefficients}) and (\ref{coefficient:e}), we finally arrive at the large-$N$ expansion of the free energy
\begin{empheq}[box=\fbox]{equation}
\begin{split}
    F&=\fft{1}{\pi\hbar\theta^3}\biggl[\fft15\left(\fft{3\pi\hbar\theta^2}{2}N\right)^\fft53\left(1-\fft{\hbar^2\theta^2}{12}+\fft{\hbar^4\theta^4}{72}+\mathcal O(\hbar^6)\right)\\
    &\kern4em+\fft{1}{12}\left(\fft{3\pi\hbar\theta^2}{2}N\right)\left(1-\fft{\hbar^2\theta^2}{3}+\fft{\hbar^4\theta^4}{12}+\mathcal O(\hbar^6)\right)\\
    &\kern4em-\fft{1}{12}\left(\fft{3\pi\hbar\theta^2}{2}N\right)^\fft23\left(1-\fft{\hbar^2\theta^2}{12}+\mathcal O(\hbar^4)\right)\\
    &\kern4em+\fft{1}{16}\left(\fft{3\pi\hbar\theta^2}{2}N\right)^\fft13\left(1-\fft{\hbar^2\theta^2(3-16\theta)}{12}+\mathcal O(\hbar^4)\right)\biggr]\\
    &\quad+\fft16\log N+\mathcal O(N^0),\label{eq:F:Fermi}
\end{split}
\end{empheq}
where we recall that $\hbar$ and $\theta$ are given in terms of the Chern-Simons levels $k_1$ and $k_2$ by (\ref{eq:theta:hbar}).  While this expression was derived for real positive $\hbar$ and $\theta$, we can analytically continue for complex parameters, provided we are careful with the fractional powers. Note that the $\mathcal O(\hbar^4\theta^4)$ terms were obtained numerically.

It is worth mentioning that one cannot reproduce the known free energy of the ABJM theory with $N^\fft32$-leading order \cite{Marino:2009jd,Drukker:2010nc,Marino:2011eh,Marino:2012az} by imposing the $\theta\to0$ limit on (\ref{eq:F:Fermi}). This is because, as we have already explained below (\ref{eq:n1(E):classic}), the final result (\ref{eq:F:Fermi}) is based on the large-$E$ expansion of the number of states under $E>1/8\theta$.

\section{Discussion}\label{sec:discussion}
When computing the free energy in the Fermi-gas approach, we have mapped the Chern-Simons levels $k_1$ and $k_2$ into $\hbar$ and $\theta$ through (\ref{eq:theta:hbar}).  This somewhat obscures the connection between $F(N;\theta,\hbar)$ in (\ref{eq:F:Fermi}) and the original parameters of the GT theory.  Inverting the relation (\ref{eq:theta:hbar}) gives
\begin{equation}
    k_1=\fft\hbar{2\pi}\left(1-\fft{i\hbar\theta}2\right)^{-1},\qquad
    k_2=-\fft\hbar{2\pi}\left(1+\fft{i\hbar\theta}2\right)^{-1}.
\end{equation}
Since the Fermi-gas computation assumed real $\hbar$ and $\theta$, the natural choice is to take $k_1$ slightly above the positive real axis and $k_2$ slightly above the negative real axis.  Of course, we expect to be able to analytically continue the result.  Nevertheless, continuation to the case where both Chern-Simons levels $k_1$ and $k_2$ are positive is a bit delicate.

In terms of $k_1$ and $k_2$, the prefactor and expansion parameters in the free energy, (\ref{eq:F:Fermi}), are given by
\begin{align}
    \fft1{\pi\hbar\theta^3}&=-2\pi i\fft{k_1-k_2}{k_1+k_2}\left(\fft{\kappa_1\kappa_2}\kappa N\right)^2,\nn\\
    \left(\fft{3\pi\hbar\theta^2}2N\right)&=\fft32\fft{k_1+k_2}{k_1-k_2}\left(\fft\kappa{\kappa_1\kappa_2}\right),\nn\\
    \hbar\theta&=-2i\fft{k_1+k_2}{k_1-k_2},
\end{align}
where we recall that $\kappa_i=k_i/N$ and $\kappa=\kappa_1+\kappa_2$.  While the semi-classical expansion corresponds to an expansion in $\hbar$, most terms in (\ref{eq:F:Fermi}) are in fact expanded in powers of $\hbar\theta$, which as we see is an expansion in inverse powers of $k_1-k_2$.  In particular, $\hbar\theta$ cannot be made small when $k_1$ and $k_2$ are both positive.

Since we have computed subleading corrections to the free energy of the GT theory using complementary approaches, it is instructive to compare both results, (\ref{eq:F:plannar:saddle}) and (\ref{eq:F:Fermi}) in the overlapping parameter regime for the two node case, $r=2$.  Perhaps the easiest way to make this comparison is to rewrite the genus zero saddle point result, (\ref{eq:GT2F0}), in terms of $\theta$ and $\hbar$ given in (\ref{eq:theta:hbar}) as
\begin{equation}
\begin{split}
	F_0&=\fft{3^\fft53\pi^\fft23}{5\cdot2^\fft53}e^{-\fft{\pi i}{6}}\left(i\hbar^2\theta\left(1+\fft{\hbar^2\theta^2}{4}\right)^{-1}\right)^\fft13N^\fft53\\
	&\quad+\fft{1}{8\theta}\left(1-\fft{\hbar^2\theta^2}{12}\right)\left(1+\fft{\hbar^2\theta^2}{4}\right)^{-1}N+\mathcal O(|\kappa|^\fft43)\\
	&=\fft1{\pi\hbar\theta^3}\biggl[\fft15\left(\fft{3\pi\hbar\theta^2N}2\right)^{\fft53}\left(1+\fft{\hbar^2\theta^2}4\right)^{-\fft13}\\
	&\kern4em+\fft1{12}\left(\fft{3\pi\hbar\theta^2 N}2\right)\left(1+\fft{\hbar^2\theta^2}{4}\right)^{-1}\left(1-\fft{\hbar^2\theta^2}{12}\right)+\cdots\biggr],
\label{eq:F:plannar:saddle:thetah}
\end{split}
\end{equation}
where we have taken
\begin{equation}
    \arg\kappa=\arg\left(\fft{i}{2\pi}\fft{\hbar^2\theta}{1+\fft{\hbar^2\theta^2}{4}}\right)\in(-\pi,-\fft{\pi}{2})\cup[0,\pi],
\end{equation}
to match the analytic continuation of the Fermi-gas free energy.  It is easy to verify that the first two terms in this expansion of the planar free energy matches the corresponding terms in the Fermi-gas result, (\ref{eq:F:Fermi}), in the overlapping parameter regime $\hbar,\theta>0$ up to $\mathcal O(\hbar^4)$.  In fact, it is this expansion that supports the extraction of the analytic coefficients in the second quantum correction to the number of states, (\ref{eq:quant2}), from the numerical computation. 

Note that the first two terms of (\ref{eq:GT2F0}) and those of (\ref{eq:F:Fermi}) are not necessarily the same as each other since they are derived using different expansion parameters presented in (\ref{expansion}). For example, if there is a genus one contribution of $\mathcal O(|\kappa|^{-\fft53}N^0)$ or $\mathcal O(|\kappa|^{-1}N^0)$ in the expansion (\ref{eq:GT2F0}), this will affect the $N^\fft53$ order or the linear $N$ order contribution in the expansion (\ref{eq:F:Fermi}) respectively. Nevertherless, the first two terms of (\ref{eq:GT2F0}) and those of (\ref{eq:F:Fermi}) match up to $\mathcal O(\hbar^6)$ corrections, which implies that the $N^\fft53$-leading and the first linear $N$ sub-leading terms of the free energy in the large-$N$ limit with fixed CS levels $k_a$ followed by the small $\hbar$ expansion comes \emph{only} from the planar limit of the free energy. Furthermore, the former seems to improve the latter by providing all order expansions of small $\hbar$ through the closed expression (\ref{eq:F:plannar:saddle:thetah}). Based on these observations, we conjecture that (\ref{eq:F:plannar:saddle:thetah}) provides the first two leading terms of the GT theory partition function precisely in the large-$N$ limit with fixed CS levels $k_a$. This is in contrast with the ABJM theory where the first $N^\fft12$ sub-leading term of the free energy receives both genus zero and genus one contributions \cite{Mezei:2013gqa}.

Curiously, except for the log term and the term with $\hbar^2\theta^2(-16\theta)$ in the penultimate line, all other terms in (\ref{eq:F:Fermi}) arise from the planar limit when matching with the 't~Hooft expansion.  In fact, the $\hbar^2\theta^2(-16\theta)$ term is rather unusual as it breaks the natural pattern of an $\hbar\theta$ expansion.  Moreover, such a term corresponds to an $\mathcal O(N^1)$ contribution in the 't~Hooft limit, which does not fit naturally into the genus expansion with contributions scaling as $\mathcal O(N^{2g-2})$.  It is not entirely clear whether this is an artifact of translating from the fixed $k_a$ expansion to the 't~Hooft limit, or if it would in the end be canceled by, \textit{e.g.}, non-perturbative terms in the Fermi-gas expansion: note that the grand potential (\ref{eq:J:1}) used to derive a partition function through (\ref{J:to:Z}) is not a simple polynomial but an infinite series for the GT theory, so the structure of non-perturbative corrections could be more involved compared to the ABJM case.  Due to its nature, this linear $N$ term is not of the universal form found in (\ref{eq:F:beyond:saddle}), and we believe in both cases that such linear $N$ terms ought not to be present in the full result.  It would be interesting to see how the fate of such terms is ultimately resolved.

In contrast with the ABJM case, where the free energy takes the form of an Airy function, there does not appear to be any nice repackaging of the large-$N$ fixed $k_a$ expansion for the free energy of the GT theory.  The origin of this difference can be traced to the form of the Wigner transformed Hamiltonian, (\ref{eq:H:Wigner:semi}), where the addition of $\fft12\theta(p+q)^2$ to the classical term, $H_W^{(0)}$, gives rise to a phase space area, $n^{(0)}(E,\theta)$, that is expanded in half-integer powers of the energy.  In particular, the perturbative expansion $\sum c_nE^{-n-\fft12}$ in (\ref{eq:n(E):classic:3}) does not terminate and does not have any obvious structure.  The resulting grand potential $J(\mu)$ then has a full perturbative expansion, as in (\ref{eq:J:1}), and hence the resulting partition function $Z$ does not have any compact expression, whether in terms of an Airy function or any other special function%
\footnote{Recall that the Airy function is a result of $J(\mu)$ being a cubic polynomial in $\mu$ up to non-perturbative terms.}.

Note that, in the Fermi-gas approach, we took as our starting point the partition function, (\ref{fermi:gas:ptn:fct}), with a position space density matrix $\rho(x_1,x_2;\theta,\Delta)$ obtained previously in \cite{Marino:2011eh}, and reviewed in Appendix~\ref{App:Fermi:setup}.  As can be seen in the appendix, this rewriting of the original matrix model partition function, (\ref{eq:Z2}), does not have a natural generalization beyond the two-node case.  Moreover, it leads to an expansion parameter, $\hbar\theta$, that is in tension with the simple picture of taking equal Chern-Simons levels for the GT theory.  To avoid both issues, we could instead start with the general density matrix for an $r$-node quiver, whose Wigner transform takes the form \cite{Marino:2011eh}
\begin{equation}
    \rho_W(q,p)=\fft1{2\cosh\fft{p}2}\star\fft1{2\cosh\fft{p-n_1q}2}\star\fft1{2\cosh\fft{p-(n_1+n_2)}2}\star\cdots\star\fft1{2\cosh\fft{p-(n_1+\cdots+n_{r-1})q}2}\star e^{\fft{i(n_1+\cdots+n_r)q^2}{2\hbar}}.
\label{eq:rhoWGT}
\end{equation}
Here we have taken Chern-Simons levels $k_a=n_ak$ and set $\hbar=2\pi k$.  The final exponential term vanishes for ABJM-like quivers, but is otherwise non-vanishing for the GT model.  The $i/\hbar$ term in the exponent is somewhat unpleasant to work with, but can be handled by a suitable analytic continuation.  (See, \textit{e.g.}, \cite{Liu:2020bih} where this was done for ABJM-like necklace quivers.)  In the two-node case, (\ref{eq:rhoWGT}) is obtained directly from (\ref{eq:Z2inter}), as in the ABJM case, without making use of the further manipulations that give rise to (\ref{fermi:gas:ptn:fct:explicit}) and (\ref{fermi:gas:ptn:fct}).  Using this expression for the density matrix gives rise to a different quantum expansion of the number of states, (\ref{eq:n(E):largeE}) and a different organization of the expansion of the free energy, (\ref{eq:F:Fermi}).  However, the full result should be identical.  It would be worthwhile to see if the use of (\ref{eq:rhoWGT}) as the starting point for the Fermi-gas approach leads to a better controlled expansion for the case when the Chern-Simons levels $k_a$ are real and have the same sign.

In the context of the AdS/CFT correspondence, the sub-leading corrections to the sphere free energy corresponds to higher derivative and quantum corrections to the holographic dual free energy of massive Type~\rom{2}A string theory with non-zero Romans mass \cite{Gaiotto:2009mv}. In light of recent developments hinting at universal features of free energies computed for higher-derivative supergravities in AdS$_4$ backgrounds \cite{Bobev:2020egg}, it would be interesting to match the sub-leading correction of the planar free energy for the GT theory in the large 't~Hooft parameter limit, namely the second term in (\ref{eq:F:plannar:saddle}), to a corresponding gravitational calculation.

One concrete place to focus on in precision holography is the study of $\log N$ corrections.  Such corrections are often well controlled in the field theory and arise from one-loop determinants on the gravity side, thus providing a probe of quantum gravity effects.  Using the Fermi-gas approach, we have demonstrated the presence of a universal $\fft16\log N$ term in the free energy, (\ref{eq:F:Fermi}), in the large-$N$ limit with fixed Chern-Simons levels $k_a$.  A similar universal $\fft14\log N$ correction has been observed in the $S^3$ free energy of ABJM theory \cite{Fuji:2011km,Marino:2011eh} and was reproduced by a one-loop contribution to the free energy in the supergravity dual \cite{Bhattacharyya:2012ye}%
\footnote{Note that some references have $-\fft14\log N$ for the log term since they take $F^{\mathrm{There}}=\log Z$ while we are using $F=-\log Z$.}.
In ABJM-like cases, the holographic dual is given by 11-dimensional supergravity on AdS$_4\times X_7$ with the tri-Sasakian manifold $X_7$ characterized by dual Chern-Simons levels $k_a$. Since the dual theory is odd-dimensional, non-zero modes do not contribute to the log term, while the zero modes yield precisely the universal $\fft14\log N$ correction.

For the GT-like Chern-Simons-matter quiver gauge theories with $\sum_ak_a\neq0$, however, the supergravity dual is not odd-dimensional.  In particular, a non-zero Romans mass dual to the sum of Chern-Simons levels does not allow for a strong coupling limit of massive Type~\rom{2}A supergravity that can lift to 11-dimensional supergravity \cite{Aharony:2010af}. Hence a one-loop calculation for the holographic dual free energy in massive Type~\rom{2}A theory will be more challenging as we would expect to obtain non-trivial contributions from the non-zero modes. Nevertheless, it would be interesting to investigate if such a one-loop calculation in Type~\rom{2}A supergravity truly yields a universal $\fft16\log N$ correction to the free energy.

\section*{Acknowledgments}

This work was supported in part by the U.S. Department of Energy under grant DE-SC0007859. JH is supported in part by a Grant for Doctoral Study from the Korea Foundation for Advanced Studies.

\appendix

\section{Free energy beyond the planar limit}\label{App:saddle:beyond}
In this Appendix, we explore the free energy $F=-\log Z$ of the GT theory beyond the planar limit using the saddle point analysis. We start from the free energy formula (\ref{eq:F:beyond}) for the two node $(r=2)$ case
\begin{equation}
	F=-N^2S_\text{cl}+2N\log N+N\log2\pi+\fft12\log\det(-\fft{\partial^2S_\text{eff}}{\partial\lambda_{a,i}\partial_{b,j}})+\mathcal O(N^0).\label{eq:F:beyond:App}
\end{equation}
The key steps here are to compute the classical action and the one-loop determinant beyond the planar limit. Since they are computed by evaluating the effective action and its second derivatives at the saddle point solution, we first need to work out the effective action (\ref{eq:Seff:continuum}) beyond the planar limit. This requires keeping track of sub-leading terms in the Euler-Maclaurin formula
\begin{equation}
	\sum_{i=m}^{n}f(i)=\int_m^nf(i)di+\fft{f(n)+f(m)}{2}+\sum_{k=1}^{\lfloor p/2\rfloor}\fft{B_{2k}}{(2k)!}(f^{(2k-1)}(n)-f^{(2k-1)}(m))+R_p,\label{Euler-Maclaurine}
\end{equation}
which have been ignored while taking the continuum limit of the effective action $S_\text{eff}$ in (\ref{eq:Seff}) to obtain (\ref{eq:Seff:continuum}). To restore these sub-leading corrections, we rewrite the Euler-Maclaurin formula (\ref{Euler-Maclaurine}) as
\begin{equation}
\begin{split}
	\sum_{i=1}^Nf(x(i))&=(N-1)\int_{-x_\ast}^{x_\ast}dx\,\rho(x)f(x)+\fft{f(x_\ast)+f(-x_\ast)}{2}\\
	&\quad+\sum_{k=1}^\infty\fft{B_{2k}}{(2k)!(N-1)^{2k-1}}\left(\fft{1}{\rho(x)}\fft{d}{dx}\right)^{2k-1}f(x)\bigg|_{-x_\ast}^{x_\ast},
\end{split}\label{Euler-Maclaurine:x}%
\end{equation}
where $x:[1,N]\to(-x_*,x_*)$ is a continuous function and $\rho(x)$ is introduced as
\begin{equation}
    di=(N-1)\rho(x)dx,\label{eq:rho:full}
\end{equation}
as in (\ref{saddle:ansatz}) and (\ref{eq:rho}). Note that the second line of (\ref{Euler-Maclaurine:x}) seems to diverge due to $\rho(\pm x_\ast)=0$. This divergence comes from na\"ively identifying $x(1)=-x_*$ and $x(N)=x_*$, and can be removed by working in the open interval. To avoid complexity, however, we keep this identification and treat the second line of (\ref{Euler-Maclaurine:x}) as sub-leading $\sum_{k=1}^\infty\mathcal O(N^{-2k+1})$ order terms in the planar limit. Namely, we ignore the ``endpoint effects.''

\subsection{Effective action beyond the planar limit}\label{App:saddle:beyond:Seff}
We now apply the full Euler-Maclaurin formula (\ref{Euler-Maclaurine:x}) to the effective action (\ref{eq:Seff}) with $r=2$ to get the continuum effective action beyond the planar limit. To begin with, the Euler-Maclaurin expansion for the first term of (\ref{eq:Seff}) with $r=2$ is given as
\begin{equation}
\begin{split}
	&\fft{i}{4\pi}\left(1-\fft1N\right)\int_{-x_\ast}^{x_\ast}dx\,\rho(x)(\kappa_1\lambda_1(x)^2+\kappa_2\lambda_2(x)^2)\\
	&+\fft{i}{8\pi N}\left[\kappa_1(\lambda_1(x_\ast)^2+\lambda_1(-x_\ast)^2)+\kappa_2(\lambda_2(x_\ast)^2+\lambda_2(-x_\ast)^2)\right]+\mathcal O(N^{-2}).\label{eq:Seff:beyond:1}
\end{split}
\end{equation}

Next, to study the Euler-Maclaurin expansion for the second term of (\ref{eq:Seff}), we introduce $\delta_i~(i=1,\cdots,N-1)$ as
\begin{equation}
    \delta_i\equiv x(i+1)-x(i)=x'(i+\epsilon_i)=\fft{1}{(N-1)\rho(x(i+\epsilon_i))}.\label{eq:delta}
\end{equation}
Here we have used the mean value theorem with some parameter $\epsilon_i\in(0,1)$ based on the continuity of $x:[1,N]\to(-x_*,x_*)$ and also used (\ref{eq:rho:full}). In the continuum limit, we can similarly define $\delta_{x\pm}$ as
\begin{equation}
	\delta_{x-}\equiv x(s)-x(s-1),\qquad\delta_{x+}\equiv x(s+1)-x(s),\label{eq:delta:continuum}
\end{equation}
where $s\in[1,N]$ such that $x(s)=x$. The Euler-Maclaurin expansion for the second term of (\ref{eq:Seff}) with $r=2$ is then given as
\begin{equation}
\begin{split}
	&(\int_{-x_\ast}^{x_\ast}\int_{-x_\ast}^{x_\ast}-\int_{-x_\ast}^{-x_\ast+\delta_1}\int_{-x_\ast}^{x}-\int_{x_\ast-\delta_{N-1}}^{x_\ast}\int_{x}^{x_\ast}-\int_{-x_\ast+\delta_1}^{x_\ast}\int_{x-\delta_{x-}}^{x}-\int_{-x_\ast}^{x_\ast-\delta_{N-1}}\int_{x}^{x+\delta_{x+}})\\
	&\quad dxdx'\,\left(1-\fft1N\right)^2\rho(x)\rho(x')\log(2\sinh\fft{\lambda_1(x)-\lambda_1(x')}{2}2\sinh\fft{\lambda_2(x)-\lambda_2(x')}{2})\\
	&+\fft{1}{N}\left(1-\fft1N\right)\int_{-x_\ast}^{x_\ast}dx\,\rho(x)\left[\log(2\sinh\fft{\lambda_1(x_\ast)-\lambda_1(x)}{2}2\sinh\fft{\lambda_2(x_\ast)-\lambda_2(x)}{2})+(x_\ast\to-x_\ast)\right]\\
	&-\fft{1}{N}\left(1-\fft1N\right)\int_{x_\ast-\delta_{N-1}}^{x_\ast}dx\,\rho(x)\log(2\sinh\fft{\lambda_1(x_\ast)-\lambda_1(x)}{2}2\sinh\fft{\lambda_2(x_\ast)-\lambda_2(x)}{2})\\
	&-\fft{1}{N}\left(1-\fft1N\right)\int_{-x_\ast}^{-x_\ast+\delta_1}dx\,\rho(x)\log(2\sinh\fft{\lambda_1(-x_\ast)-\lambda_1(x)}{2}2\sinh\fft{\lambda_2(-x_\ast)-\lambda_2(x)}{2})\\
	&+\fft{1}{2N}\left(1-\fft1N\right)\int_{-x_\ast+\delta_1}^{x_\ast}dx\,\rho(x)\log(2\sinh\fft{\lambda_1(x)-\lambda_1(x-\delta_{x-})}{2}2\sinh\fft{\lambda_2(x)-\lambda_2(x-\delta_{x-})}{2})\\
	&+\fft{1}{2N}\left(1-\fft1N\right)\int_{-x_\ast}^{x_\ast-\delta_{N-1}}dx\,\rho(x)\log(2\sinh\fft{\lambda_1(x)-\lambda_1(x+\delta_{x+})}{2}2\sinh\fft{\lambda_2(x)-\lambda_2(x+\delta_{x+})}{2})\\
	&+\fft{3}{4N^2}\log(2\sinh\fft{\lambda_1(x_\ast-\delta_{N-1})-\lambda_1(x_\ast)}{2}2\sinh\fft{\lambda_2(x_\ast-\delta_{N-1})-\lambda_2(x_\ast)}{2})\\
	&+\fft{3}{4N^2}\log(2\sinh\fft{\lambda_1(-x_\ast)-\lambda_1(-x_\ast+\delta_1)}{2}2\sinh\fft{\lambda_2(-x_\ast)-\lambda_2(-x_\ast+\delta_1)}{2})\\
	&-\fft4N\sum_{k=1}^\infty\fft{B_{2k}}{2k(2k-1)}+\mathcal O(N^{-2}).\label{eq:Seff:beyond:2}
\end{split}
\end{equation}

Lastly, the Euler-Maclaurin expansion for the third term of (\ref{eq:Seff}) with $r=2$ is given as
\begin{equation}
\begin{split}
	&-2\left(1-\fft1N\right)^2\int_{-x_\ast}^{x_\ast}dx\,\rho(x)\int_{-x_\ast}^{x_\ast}dx'\,\rho(x')\log(2\cosh\fft{\lambda_1(x)-\lambda_2(x')}{2})\\
	&-\fft{1}{N}\left(1-\fft1N\right)\int_{-x_\ast}^{x_\ast}dx\,\rho(x)\left[\log(2\cosh\fft{\lambda_1(x)-\lambda_2(x_\ast)}{2}2\cosh\fft{\lambda_1(x)-\lambda_2(-x_\ast)}{2})+(\lambda_1\leftrightarrow\lambda_2)\right]\\
	&+\mathcal O(N^{-2}).\label{eq:Seff:beyond:3}
\end{split}
\end{equation}

The total effective action beyond the planar limit is the sum of all three contributions, (\ref{eq:Seff:beyond:1}), (\ref{eq:Seff:beyond:2}), and (\ref{eq:Seff:beyond:3}).

\subsection{Classical action beyond the planar limit}\label{App:saddle:beyond:Scl}
The classical action beyond the planar limit can be obtained by evaluating the effective action computed in the previous subsection \ref{App:saddle:beyond:Seff} at the saddle point. For simplicity, we use the $|\kappa|^\alpha$-leading order saddle point solution
\begin{subequations}
\begin{align}
    \lambda_a(x)&=|\kappa|^{-\alpha}(x+iy(x)),\\
	y(x)&=mx\quad\text{where }\quad\fft{(1+im)^3}{(1+m^2)^\fft32}=i\fft{|\kappa|}{\kappa},\\
	\rho(x)&=\fft{(1+m^2)^\fft12}{8\pi}\left(6^\fft23-\fft{1+m^2}{\pi^2}x^2\right),
\end{align}\label{sol:saddle:beyond}%
\end{subequations}
which is from (\ref{saddle:ansatz:smallk}), (\ref{eq:rho:smallk}), and (\ref{sol:saddle:1}) in the main text with $r=2$. 

Substituting the saddle point solution (\ref{sol:saddle:beyond}) into the three contributions to the effective action, (\ref{eq:Seff:beyond:1}), (\ref{eq:Seff:beyond:2}), and (\ref{eq:Seff:beyond:3}) gives the following results:

\noindent\textbf{1st term}
\begin{equation}
\begin{split}
	&\fft{i}{4\pi}\left(1-\fft1N\right)\kappa|\kappa|^{-2\alpha}(1+im)^2\int_{-x_\ast}^{x_\ast}dx\,\rho(x)x^2+\fft{i}{4\pi N}\kappa|\kappa|^{-2\alpha}(1+im)^2x_\ast^2+\mathcal O(N^{-2}).\label{eq:Scl:beyond:1}
\end{split}
\end{equation}

\noindent\textbf{2nd term}
\begin{equation}
\begin{split}
	&(\int_{-x_\ast}^{x_\ast}\int_{-x_\ast}^{x_\ast}-2\int_{-x_\ast}^{-x_\ast+\delta_1}\int_{-x_\ast}^{x}-2\int_{-x_\ast+\delta_1}^{x_\ast}\int_{x-\delta_{x-}}^{x})\\
	&\quad dxdx'\,\left(1-\fft1N\right)^2\rho(x)\rho(x')\log(4\sinh^2\fft{|\kappa|^{-\alpha}(1+im)(x-x')}{2})\\
	&+\fft{2}{N}\left(1-\fft1N\right)\int_{-x_\ast}^{x_\ast}dx\,\rho(x)\log(4\sinh^2\fft{|\kappa|^{-\alpha}(1+im)(x-x_\ast)}{2})\\
	&-\fft{2}{N}\left(1-\fft1N\right)\int_{-x_\ast}^{-x_\ast+\delta_1}dx\,\rho(x)\log(4\sinh^2\fft{|\kappa|^{-\alpha}(1+im)(x+x_\ast)}{2})\\
	&+\fft{1}{N}\left(1-\fft1N\right)\int_{-x_\ast+\delta_1}^{x_\ast}dx\,\rho(x)\log(4\sinh^2\fft{|\kappa|^{-\alpha}(1+im)\delta_{x-}}{2})\\
	&+\fft{3}{2N^2}\log(4\sinh^2\fft{|\kappa|^{-\alpha}(1+im)\delta_1}{2})-\fft4N\sum_{k=1}^\infty\fft{B_{2k}}{2k(2k-1)}+\mathcal O(N^{-2}).\label{eq:Scl:beyond:2}
\end{split}
\end{equation}

\noindent\textbf{3rd term}
\begin{equation}
\begin{split}
    &-\left(1-\fft1N\right)^2\int_{-x_\ast}^{x_\ast}dx\,\rho(x)\int_{-x_\ast}^{x_\ast}dx'\,\rho(x')\log(4\cosh^2\fft{|\kappa|^{-\alpha}(1+im)(x-x')}{2})\\
	&-\fft{2}{N}\left(1-\fft1N\right)\int_{-x_\ast}^{x_\ast}dx\,\rho(x)\log(4\cosh^2\fft{|\kappa|^{-\alpha}(1+im)(x-x_\ast)}{2})+\mathcal O(N^{-2}).\label{eq:Scl:beyond:3}
\end{split}
\end{equation}

\noindent For the 2nd term, we have used that $\rho(x)$ in (\ref{sol:saddle:beyond}) is an even function of $x$ and therefore $\delta_1=\delta_{N-1}$ and $\delta_{x\pm}=\delta_{(-x)\mp}$.

Note that we have not yet substituted actual values of $\rho(x)$ given in (\ref{sol:saddle:beyond}) in order to keep the expressions compact. Summing over all three contributions, (\ref{eq:Scl:beyond:1}), (\ref{eq:Scl:beyond:2}), and (\ref{eq:Scl:beyond:3}), we obtain the classical action $S_\text{cl}$ beyond the planar limit
\begin{equation}
    S_\text{cl}=S_\text{cl,planar}+S_\text{cl,non-planar},
\label{eq:Scl:split}
\end{equation}
where we have distinguished the planar part and the non-planar part as
\begin{equation}
    S_\text{cl,planar}=|\kappa|^\alpha\left(\fft{i}{4\pi}\fft{\kappa}{|\kappa|}(1+im)^2\int_{-x_\ast}^{x_\ast}dx\,\rho(x)x^2-\fft{\pi^2}{1+im}\int_{-x_\ast}^{x_\ast}dx\,\rho(x)^2\right)
\end{equation}
and
\begin{equation}
\begin{split}
	S_\text{cl,non-planar}&=\fft{i}{4\pi N}\kappa|\kappa|^{-2\alpha}(1+im)^2\left(x_\ast^2-\int_{-x_\ast}^{x_\ast}dx\,\rho(x)x^2\right)\\
	&\quad-\fft1N\left(2-\fft1N\right)\int_{-x_\ast}^{x_\ast}dx\,\rho(x)\int_{-x_\ast}^{x_\ast}dx'\,\rho(x')\log(\tanh^2\fft{|\kappa|^{-\alpha}(1+im)(x-x')}{2})\\
	&\quad-2\left(1-\fft1N\right)^2\int_{-x_\ast}^{-x_\ast+\delta_1}dx\,\rho(x)\int_{-x_\ast}^{x}dx'\rho(x')\log(4\sinh^2\fft{|\kappa|^{-\alpha}(1+im)(x-x')}{2})\\
	&\quad-2\left(1-\fft1N\right)^2\int_{-x_\ast+\delta_1}^{x_\ast}dx\,\rho(x)\int_{x-\delta_{x-}}^{x}dx'\,\rho(x')\log(4\sinh^2\fft{|\kappa|^{-\alpha}(1+im)(x-x')}{2})\\
	&\quad+\fft2N\left(1-\fft1N\right)\int_{-x_\ast}^{x_\ast}dx\,\rho(x)\log(\tanh^2\fft{|\kappa|^{-\alpha}(1+im)(x+x_\ast)}{2})\\
	&\quad-\fft2N\left(1-\fft1N\right)\int_{-x_\ast}^{-x_\ast+\delta_1}dx\,\rho(x)\log(4\sinh^2\fft{|\kappa|^{-\alpha}(1+im)(x+x_\ast)}{2})\\
	&\quad+\fft{1}{N}\left(1-\fft1N\right)\int_{-x_\ast+\delta_1}^{x_\ast}dx\,\rho(x)\log(4\sinh^2\fft{|\kappa|^{-\alpha}(1+im)\delta_{x-}}{2})\\
	&\quad+\fft{3}{2N^2}\log(4\sinh^2\fft{|\kappa|^{-\alpha}(1+im)\delta_1}{2})\\
	&\quad-\fft4N\sum_{k=1}^\infty\fft{B_{2k}}{2k(2k-1)}+\mathcal O(N^{-2}).
\end{split}\label{eq:Scl:beyond}
\end{equation}
Note that, while the above results go beyond the planar limit, they are only valid to leading order in the small-$|\kappa|$ expansion since we have used the $|\kappa|^\alpha$-leading order saddle point solution (\ref{sol:saddle:beyond}). Hence the planar part $S_\text{cl,planar}$ matches the $|\kappa|^\alpha$-leading order classical action (\ref{eq:Seff:1:withend}) as expected, but does not include the subleading order $S_3$ in (\ref{eq:Seff:3:simple}).

Now we evaluate the integrals in the non-planar corrections to the classical action (\ref{eq:Scl:beyond}). Using (\ref{eq:delta}) and (\ref{eq:delta:continuum}), and ignoring endpoint effects as mentioned above, we find
\begin{subequations}
\begin{align}
    S_\text{cl,non-planar}^\text{4th line}&=\fft{4}{N}\left(1+\log\fft{N}{|\kappa|^{-\alpha}(1+im)}\right)\int_{-x_\ast+\delta_1}^{x_\ast}dx\,\rho(x)\nn\\
	&\quad+\fft{4}{N}\int_{-x_\ast+\delta_1}^{x_\ast}dx\,\rho(x)\log\rho(x)+\mathcal O(N^{-2}\log N),\\
    S_\text{cl,non-planar}^\text{7th line}&=-\fft2N\log\fft{N}{|\kappa|^{-\alpha}(1+im)}\int_{-x_\ast+\delta_1}^{x_\ast}dx\,\rho(x)\nn\\
    &\quad-\fft2N\int_{-x_\ast+\delta_1}^{x_\ast}dx\,\rho(x)\log\rho(x)+\mathcal O(N^{-2}),\\
    S_\text{cl,non-planar}^\text{9th line}&=-4\int_0^\infty dt\,t^{-2}e^{-t}\left(\fft{t}{e^t-1}-1+\fft12t\right)=-\fft1N(4-2\log2\pi),\label{eq:Scl:beyond:evaluation:9}
\end{align}\label{eq:Scl:beyond:evaluation}%
\end{subequations}
where the other terms are of $\mathcal O(N^{-2}\log N)$ order at most. The last expression, (\ref{eq:Scl:beyond:evaluation:9}), is obtained by substituting $z=1$ into the integral representation of the gamma function
\begin{equation}
	\log\Gamma(z)=(z-\fft12)\log z-z+\fft12\log2\pi+\int_0^\infty dt\,t^{-2}e^{-zt}\left(\fft{t}{e^t-1}-1+\fft12t\right).
\end{equation}
Using actual values of $\rho(x)$ given in (\ref{sol:saddle:beyond}), we can evaluate the integrals in (\ref{eq:Scl:beyond:evaluation}) as
\begin{subequations}
\begin{align}
	\int_{-x_\ast+\delta_1}^{x_\ast}dx\,\rho(x)&=1+\mathcal O(\delta_1^2),\\
	\int_{-x_\ast+\delta_1}^{x_\ast}dx\,\rho(x)\log\rho(x)&=\fft16\left(-10+\log(\fft{81(1+m^2)^3}{4\pi^6})\right)+\mathcal O(\delta_1^2\log\delta_1).
\end{align}\label{rho:integral}%
\end{subequations}
Finally, substituting (\ref{rho:integral}) into (\ref{eq:Scl:beyond:evaluation}) then (\ref{eq:Scl:beyond}), we get the contribution from the classical action to the free energy in (\ref{eq:F:beyond:App}) beyond the planar limit
\begin{empheq}[box=\fbox]{equation}
\begin{split}
	-N^2S_\text{cl,non-planar}&=-2N\log N-2N\left(\log2\pi-\log(|\kappa|^{-\alpha}(1+im))\right)\\
	&\quad-\fft{N}{3}\left(-10+\log(\fft{81(1+m^2)^3}{4\pi^6})\right)+\mathcal O(\log N).\label{eq:Scl:beyond:simple}
\end{split}
\end{empheq}
%

\subsection{One-loop determinant beyond the planar limit}
To evaluate the contribution from the one-loop determinant in (\ref{eq:F:beyond:App}), first we write down the components of the Hessian matrix explicitly as 
\begin{subequations}
\begin{align}
	-\fft{\partial^2S_\text{eff}}{\partial\lambda_{1,i}\partial\lambda_{1,j}}&=\delta_{ij}\left(-\fft{i}{2\pi N}\kappa_1+\fft{1}{2N^2}\sum_{k=1\,(\neq i)}^N\csch^2\fft{\lambda_{1,i}-\lambda_{1,k}}{2}+\fft{1}{2N^2}\sum_{k=1}^N\sech^2\fft{\lambda_{1,i}-\lambda_{2,k}}{2}\right)\nn\\
	&\quad-(1-\delta_{ij})\fft{1}{2N^2}\csch^2\fft{\lambda_{1,i}-\lambda_{1,j}}{2},\\
	-\fft{\partial^2S_\text{eff}}{\partial\lambda_{1,i}\partial\lambda_{2,j}}&=-\fft{1}{2N^2}\sech^2\fft{\lambda_{1,i}-\lambda_{2,j}}{2},\\
	-\fft{\partial^2S_\text{eff}}{\partial\lambda_{2,i}\partial\lambda_{2,j}}&=\delta_{ij}\left(-\fft{i}{2\pi N}\kappa_2+\fft{1}{2N^2}\sum_{k=1\,(\neq i)}^N\csch^2\fft{\lambda_{2,i}-\lambda_{2,k}}{2}+\fft{1}{2N^2}\sum_{k=1}^N\sech^2\fft{\lambda_{2,i}-\lambda_{1,k}}{2}\right)\nn\\
	&\quad-(1-\delta_{ij})\fft{1}{2N^2}\csch^2\fft{\lambda_{2,i}-\lambda_{2,j}}{2}.
\end{align}\label{Hessian:components}%
\end{subequations}
Once again, we need to replace the sums by integrals using the Euler-Maclaurin formula, (\ref{Euler-Maclaurine:x}). In principle, the result would be equivalent to taking the second order functional derivatives of the continued effective action give above with respect to the eigenvalue distribution $\lambda_a(x)$, namely $-{\delta^2S_\text{eff}}/{\delta\lambda_a(x(i))\delta\lambda_b(x(j))}$. However, we find it simpler to work directly with (\ref{Hessian:components}).

To estimate its determinant, we split the Hessian matrix into its diagonal and off-diagonal components
\begin{equation}
	-\fft{\partial^2S_\text{eff}}{\partial\lambda_{a,i}\partial\lambda_{b,j}}=\delta_{ab}\delta_{ij}A_{ai}+(1-\delta_{ab}\delta_{ij})B_{ai,bj}.
\end{equation}
Then the one-loop determinant contribution to the free energy given in (\ref{eq:F:beyond:App}) can be rewritten as
\begin{equation}
	F\,\supseteq~\fft12\log\det(-\fft{\partial^2S_\text{eff}}{\partial\lambda_{a,i}\partial\lambda_{b,j}})=\fft12\log\det A+\fft12\log\det(I+A^{-1}B).\label{one-loop:split}
\end{equation}
We investigate the diagonal part and the off-diagonal part in order.

\subsubsection*{Diagonal part}
The contribution from the diagonal part in (\ref{one-loop:split}) is given explicitly from the matrix components (\ref{Hessian:components}) as
\begin{equation}
    \fft12\log\det A=\fft12\sum_{i=1}^N\left(\log(-\fft{\partial^2S}{\partial\lambda_{1,i}^2})+\log(-\fft{\partial^2S}{\partial\lambda_{2,i}^2})\right).\label{Diagonal}
\end{equation}
We would like to continue this diagonal contribution using the Euler-Maclaurin expansion (\ref{Euler-Maclaurine:x}). To begin with, we derive the Euler-Maclaurine expansion of $\fft{\partial^2S}{\partial\lambda_i^2}$ as
\begin{equation}
\begin{split}
	-\fft{\partial^2S}{\partial\lambda_i^2}&=\fft{1}{2N}\left[\left(1-\fft1N\right)(\int_{-x_\ast}^{x_i-\delta_i}+\int_{x_i+\delta_{i+1}}^{x_\ast})dx'\,\rho(x')\csch^2\fft{\lambda_i-\lambda(x')}{2}\right.\\
	&\kern3em~+\fft{1}{2N}(\csch^2\fft{\lambda_i-\lambda(x_i-\delta_i)}{2}+\csch^2\fft{\lambda_i-\lambda(-x_\ast)}{2}\\
	&\kern3em~+\csch^2\fft{\lambda_i-\lambda(x_\ast)}{2}+\csch^2\fft{\lambda_i-\lambda(x_i+\delta_{i+1})}{2})\\
	&\kern3em~\left.+2\sum_{k=1}^{\infty}\fft{B_{2k}}{(2k)!N(N-1)^{2k-1}}\fft{1}{\rho(x_i)^{2k-1}}\fft{4(2k)!}{|\kappa|^{-2\alpha}(1+im)^2\delta_i^{2k+1}}\right]\\
	&\quad+\fft{1}{2N}\left[\left(1-\fft1N\right)\int_{-x_\ast}^{x_\ast}dx'\,\rho(x')\sech^2\fft{\lambda_i-\tilde\lambda(x')}{2}\right.\\
	&\kern4em\left.+\fft{1}{2N}(\sech^2\fft{\lambda_i-\tilde\lambda(x_\ast)}{2}+\sech^2\fft{\lambda_i-\tilde\lambda(-x_\ast)}{2})\right]+\mathcal O(N^{-1}).
\end{split}
\end{equation}
Evaluating the above expression at the saddle point solution (\ref{sol:saddle:beyond}) with $\delta_i,\delta_{i+1}\sim\fft{1}{N\rho(x_i)}$ from (\ref{eq:delta}) then gives
\begin{equation}
\begin{split}
	-\fft{\partial^2S}{\partial\lambda_i^2}&=\fft{1}{2N}\left[\fft{4\rho(x_i)}{|\kappa|^{-2\alpha}(1+im)^2\delta_i}+\fft{4\rho(x_i)}{|\kappa|^{-2\alpha}(1+im)^2\delta_{i+1}}+\mathcal O(\delta^0)\right.\\
	&\kern3em~+\fft{1}{2N}\left(\fft{4}{|\kappa|^{-2\alpha}(1+im)^2\delta_i^2}+\fft{4}{|\kappa|^{-2\alpha}(1+im)^2\delta_{i+1}^2}+\mathcal O(\delta^0)\right)\\
	&\kern3em~\left.\fft{8N\rho(x_i)^2}{|\kappa|^{-2\alpha}(1+im)^2}\sum_{k=1}^\infty B_{2k}\right]+\mathcal O(N^{-1})\\
	&=\fft{2\pi^2\rho(x_i)^2}{3|\kappa|^{-2\alpha}(1+im)^2}+\mathcal O(N^{-1}).
\end{split}\label{S:2deriv}
\end{equation}
In the second equation of (\ref{S:2deriv}), we have used the identity $(B_0=1,B_1=-\fft12)$
\begin{equation}
	\sum_{k=1}^\infty B_{2k}=\int_0^\infty dt\,\fft{te^{-t}}{e^t-1}-B_0-B_1=\fft{\pi^2}{6}-\fft32.
\end{equation}
Substituting (\ref{S:2deriv}) into (\ref{Diagonal}) and using $-\fft{\partial^2S}{\partial\lambda_i^2}=-\fft{\partial^2S}{\partial\tilde\lambda_i^2}$ at the saddle point solution (\ref{sol:saddle:beyond}) then gives
\begin{equation}
    \fft12\log\det A=\sum_{i=1}^N\log(\fft{2\pi^2\rho(x_i)^2}{3|\kappa|^{-2\alpha}(1+im)^2})+\mathcal O(N^0).
\end{equation}
Finally, applying the Euler-Maclaurin formula (\ref{Euler-Maclaurine:x}) one more time and using the second integral of (\ref{rho:integral}), we get
\begin{empheq}[box=\fbox]{equation}
	\fft12\log\det A=-N\log\fft{3|\kappa|^{-2\alpha}(1+im)^2}{2\pi^2}+\fft{N}{3}\left(-10+\log(\fft{81(1+im)^3}{4\pi^6})\right)+o(N).\label{Diagonal:simple}
\end{empheq}
%

\subsubsection*{Off-diagonal part}
The contribution from the off-diagonal part in (\ref{one-loop:split}) is given explicitly from the matrix components (\ref{Hessian:components}) as
\begin{equation}
\begin{split}
	\fft12\Tr\log(I+A^{-1}B)&=-\fft12\sum_{n=1}^\infty\fft1n\Tr(X^n)\\
	&=-\fft12\sum_{n=1}^\infty\fft1n\sum_{I_1,\cdots,I_n=1}^{2N}X_{I_1I_2}X_{I_2I_3}\cdots X_{I_nI_1},
\end{split}\label{Off-diagonal}
\end{equation}
where we have defined a $2N\times 2N$ matrix $X$ as $(i,j=1,\cdots,N)$
\begin{equation}
\begin{split}
	X&\equiv-A^{-1}B\\
	&=\fft{3|\kappa|^{-2\alpha}(1+im)^2}{4\pi^2N^2}\begin{pmatrix}
	(1-\delta_{ij})\fft{\csch^2\fft{\lambda_{1,i}-\lambda_{1,j}}{2}}{\rho(x_i)^2} & \fft{\sech^2\fft{\lambda_{1,i}-\lambda_{2,j}}{2}}{\rho(x_i)^2}\\
	\fft{\sech^2\fft{\lambda_{2,i}-\lambda_{1,j}}{2}}{\rho(x_i)^2} & (1-\delta_{ij})\fft{\csch^2\fft{\lambda_{2,i}-\lambda_{2,j}}{2}}{\rho(x_i)^2}
	\end{pmatrix}.
\end{split}
\end{equation}

It is highly involved to compute $\Tr(X^n)$ completely: so we focus on its linear $N$-leading order contribution to the one-loop determinant through (\ref{Off-diagonal}). Due to the apparent $N^{-2}$ scale of the matrix $X$, the linear $N$-leading order contribution of $\Tr(X^n)$ would come from the multiplications of $\csch^2\fft{\lambda_{a,i}-\lambda_{a,j}}{2}$ terms with nearby $i,j$ that possibly cancel the $N^{-2}$ factor. As a result, we can keep track of the linear $N$-leading order of the one-loop determinant (\ref{Off-diagonal}) as
\begin{equation}
	\fft12\Tr\log(I+A^{-1}B)=-\sum_{n=1}^\infty\fft1nI_n+o(N),\label{Off-diagonal:simple}
\end{equation}
where we have derived $I_n$ as
\begin{equation}
\begin{split}
    I_n&=\begin{cases}
	0 & (n=1);\\
	\left(\fft{3}{\pi^2N^2}\right)^n\sum_{i_1,\cdots i_n=1}^{\prime\,N}\fft{1}{\rho(x_{i_1})^2\cdots\rho(x_{i_n})^2}\fft{1}{(x_{i_1}-x_{i_2})^2\cdots(x_{i_n}-x_{i_1})^2} & (n\geq2),
	\end{cases}\label{eq:I_n}
\end{split}
\end{equation}
using $\lambda_i=|\kappa|^{-\alpha}(1+im)x_i$ from the saddle point solution in the small-$|\kappa|$ limit (\ref{sol:saddle:beyond}). The primed sum in (\ref{eq:I_n}) means that we have excluded the cases with $i_k=i_{k+1}~(i_{n+1}\equiv i_1)$ for any $k\in\{1,\cdots,n\}$ to avoid a vanishing denominator.

Now the goal is to figure out the linear $N$-leading order of $I_n$ in (\ref{eq:I_n}). For that purpose, we simplify (\ref{eq:I_n}) further by rewriting $\rho(x_i)$ as 
\begin{equation}
	\rho(x_i)^2\sim\fft{1}{N^2(x_i-x_{i+1})^2}\sim\fft{(i-j)^2}{N^2(x_i-x_j)^2}\label{eq:rho(x_i)}
\end{equation}
using (\ref{eq:delta}), provided $x_i,x_j$ are not close to the endpoints and $|i-j|=\mathcal O(N^0)$. This assumption is consistent with the fact that we are ignoring endpoint effects. Substituting (\ref{eq:rho(x_i)}) into (\ref{eq:I_n}) then gives
\begin{equation}
\begin{split}
	I_n&=\left(\fft{3}{\pi^2}\right)^n\sideset{}{'}\sum_{i_1,\cdots, i_n=1}^N\fft{1}{(i_1-i_2)^2\cdots (i_{n-1}-i_n)^2(i_n-i_1)^2}+o(N)\\
	&=N\left(\fft{3}{\pi^2}\right)^n\sideset{}{'}\sum_{i_2,\cdots, i_n}\fft{1}{i_2^2(i_2-i_3)^2\cdots (i_{n-1}-i_n)^2i_n^2}+o(N)\\
	&=N\left(\fft{3}{\pi^2}\right)^n\sideset{}{'}\sum_{i_1,\cdots, i_{n-1}}\fft{1}{i_1^2i_2^2\cdots i_{n-1}^2(i_1+i_2+\cdots+i_{n-1})^2}+o(N).
\end{split}\label{eq:I_n:simple}%
\end{equation}
The second equation is derived under the large-$N$ limit considering the values of $i_1$ as a flat direction. The third equation is obtained from a simple renaming of indices. The primed sums in the second and third lines represent a sum over all integers except the ones that make the denominator vanish.

To compute the above infinite series and thereby determine the linear $N$-leading order of $I_n$ in (\ref{eq:I_n:simple}), we introduce $G_{a,n}~(a=1,2,\cdots,\,n=0,1,\cdots)$ as
\begin{subequations}
\begin{align}
	G_{a,n}&\equiv\left(\fft{3}{\pi^2}\right)^n\sum_{i_1,\cdots i_{n-1}}{}'\fft{1}{i_1^2i_2^2\cdots i_{n-1}^2(i_1+i_2+\cdots+i_{n-1})^{2a}},\\
	I_n&=G_{1,n}N+o(N).\label{eq:I_n:leading}
\end{align}
\end{subequations}
Here we set $G_{a,0}=1$ and $G_{a,1}=0$. What we want after all would be $G_{1,n}$, namely the linear $N$-leading order of $I_n$. To obtain $G_{1,n}$, however, we will need to investigate a generalized series $G_{a,n}$ first.

To begin with, note that we can compute $G_{a,2}$ directly as
\begin{equation}
	G_{a,2}=\fft{18\zeta(2(a+1))}{\pi^4}.
\end{equation}
General $G_{a,n}$ is then determined by the initial values $G_{a,0}$, $G_{a,1}$, and $G_{a,2}$ and the recurrence relation
\begin{equation}
\begin{split}
	G_{a,n}&=\left(\fft{3}{\pi^2}\right)^n\left[\sideset{}{'}\sum_{i_1,\cdots,i_{n-2}}^{i_1+\cdots+i_{n-2}\neq0}\fft{1}{i_1^2\cdots i_{n-2}^2}\sideset{}{'}\sum_{i_{n-1}}\fft{1}{i_{n-1}^2(i_1+\cdots i_{n-1})^{2a}}\right.\\
	&\kern5em\left.+\sideset{}{'}\sum_{i_1,\cdots,i_{n-3}}\fft{1}{i_1^2\cdots i_{n-3}^2(i_1+\cdots i_{n-3})^2}\sideset{}{'}\sum_{i_{n-1}}\fft{1}{i_{n-1}^{2(a+1)}}\right]\\
	&=\fft{18\zeta(2(a+1))}{\pi^4}G_{1,n-2}+2aG_{a,n-1}-\fft{3(a+1)(2a+1)}{\pi^2}G_{a+1,n-1}\\
	&\quad+\fft{6}{\pi^2}\sum_{k=1}^{a-1}(2k-1)\zeta(2(a-k+1))G_{k,n-1},\label{recurrence:G}
\end{split}
\end{equation}
where the second equation has been derived using the identity
\begin{equation}
\begin{split}
	&\sideset{}{'}\sum_{i}^{j\neq0}\fft{1}{i^2(i+j)^{2a}}=2\sum_{k=1}^{a-1}\fft{(2k-1)\zeta(2(a-k+1))}{j^{2k}}+\fft{2\pi^2a}{3j^{2a}}-\fft{(a+1)(2a+1)}{j^{2(a+1)}}.
\end{split}
\end{equation}
Starting from the recurrence relation of $G_{a,n}$ (\ref{recurrence:G}) and using the identities
\begin{subequations}
\begin{align}
	\sum_{k=1}^a\zeta(2k)\zeta(2(a+1-k))&=\fft{2a+3}{2}\zeta(2(a+1)),\\
	\sum_{k=1}^ak\zeta(2k)\zeta(2(a+1-k))&=\fft{(a+1)(2a+3)}{4}\zeta(2(a+1)),
\end{align}
\end{subequations}
we deduce the recurrence relation of $G_{1,n}$ as
\begin{equation}
	G_{1,n+2}=2G_{1,n+1}-G_{1,n}+\fft{18}{\pi^2}H_{2,n}\quad(n\geq1),\label{recurrence:G1}
\end{equation}
with the initial values $G_{1,0}=1$, $G_{1,1}=0$, and $G_{1,2}=\fft15$. Here $H_{a,n}$ is defined as
\begin{equation}
	H_{a,n}\equiv\fft{3}{\pi^2}\left(\fft{6}{\pi^2}\right)^n\sum_{k=0}^nb_{n,k}\zeta(2(a+k))\zeta(2(n-k))\quad(a\geq2,~n\geq1).
\end{equation}
Here the coefficients $b_{n,k}$ are determined by the recurrence relation
\begin{equation}
\begin{split}
	b_{n+1,k}&=-\fft{k(2k-1)}{2}b_{n,k-1}+2k\fft{\zeta(2)\zeta(2(n-k))}{\zeta(2(n+1-k))}b_{n,k}\\
	&\quad+(2k-1)\sum_{l=2}^{n+1-k}\fft{\zeta(2l)\zeta(2(n+1-k-l))}{\zeta(2(n+1-k))}b_{n,k+l-1}\quad(k=0,1,\ldots, n+1)
\end{split}
\end{equation}
with the initial values $b_{1,0}=0$, $b_{1,1}=2$, and the convention $b_{n,-\mathbb N}=0$. Using the recurrence relation (\ref{recurrence:G1}), we have computed $G_{1,n}$ for $n$ up to $300$ and the first few of them are given as
\begin{equation}
	G_{1,3}=\fft{2}{35},\quad G_{1,4}=\fft{3}{35},\quad G_{1,5}=\fft{4}{77},\quad G_{1,6}=\fft{53}{1001},\quad\ldots.\label{series:G1}
\end{equation}

\begin{figure}[t]
	\centering
	\includegraphics[scale=.65]{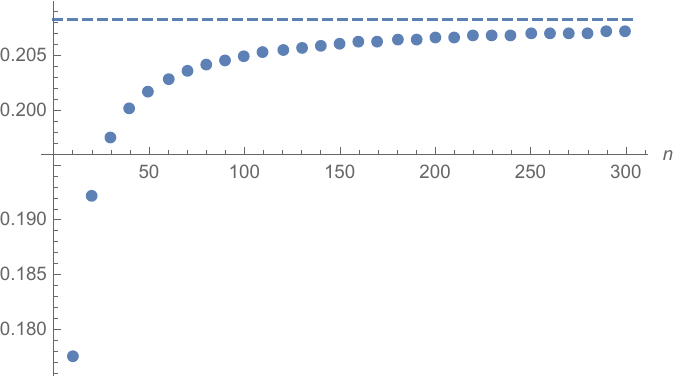}
	\caption{Dots represent partial sums $\sum_{k=1}^n\fft{G_{1,k}}{k}$ with $n=10,20,\ldots,300$. The dashed line represents the numerical value of $2-\log6$. \label{1-loop:off-diagonal}}
\end{figure}

Combining (\ref{Off-diagonal:simple}) and (\ref{eq:I_n:leading}) gives the linear $N$-leading order of the one-loop determinant in terms of the series $G_{1,n}$ obtained in (\ref{series:G1}) as
\begin{empheq}[box=\fbox]{equation}
    \fft12\log\Tr\log(I+A^{-1}B)=-N\sum_{n=1}^\infty\fft{G_{1,n}}{n}+o(N)\overset{!}{=}-(2-\log 6)N+o(N).\label{Off-diagonal:numeric}
\end{empheq}
We could not analytically prove the above numerical identity. See Figure~\ref{1-loop:off-diagonal} for numerical evidence.

Finally, substituting the diagonal contribution (\ref{Diagonal:simple}) and the off-diagonal contribution (\ref{Off-diagonal:numeric}) into (\ref{one-loop:split}) gives the one-loop determinant contribution to the free energy (\ref{eq:F:beyond:App}) beyond the planar limit
\begin{empheq}[box=\fbox]{equation}
\begin{split}
	\fft12\log\det(-\fft{\partial^2S_\text{eff}}{\partial\lambda_{a,i}\partial\lambda_{b,j}})&=-2N+2N(\log2\pi-\log|\kappa|^{-\alpha}(1+im))\\
	&\quad+\fft{N}{3}\left(-10+\fft{81(1+im)^3}{4\pi^6}\right)+o(N).\label{eq:one-loop:beyond:simple}
\end{split}
\end{empheq}
%

\section{Conventions and integrals used in the ideal Fermi-gas approach}\label{App:Fermi}

\subsection{Trace of a function of operators}
\label{App:Fermi:trace}
In this Appendix we review how to compute the trace of a function of operators in quantum mechanics with conventions
\begin{subequations}
	\begin{align}
	[\hat x,\hat p]&=i\hbar,\\
	\hat x|x\rangle&=x|x\rangle,\\
	\hat p|p\rangle&=p|p\rangle,\\
	\qquad\langle x|x'\rangle&=\delta(x-x'),\\
	\qquad \langle p|p'\rangle&=\delta(p-p'),\\
	1&=\int dx |x\rangle\langle x|=\int dp |p\rangle\langle p|,\\
	\langle x|p\rangle&=\fft{1}{\sqrt{2\pi\hbar}}e^{ipx/\hbar}.
	\end{align}
\end{subequations}
First, we introduce the Wigner transformation of an operator $\hat A$, namely $A_W(q,p)$, as
\begin{equation}
	A_W(q,p)\equiv\int dq'\langle q-\fft{q'}{2}|\hat A|q+\fft{q'}{2}\rangle e^{ipq'/\hbar}.\label{Wigner}
\end{equation}
In terms of the Wigner transformation, we can write the trace of an operator $\hat A$ as
\begin{equation}
	\Tr\hat A=\int\fft{dpdq}{2\pi\hbar}A_W(q,p).\label{Wigner:trace}
\end{equation}
It is also well known that the Wigner transform of a product of operators is given as (see, \textit{e.g.}, (3.36) and (3.37) of \cite{Marino:2011eh})
\begin{equation}
\begin{split}
	(\hat A\hat B)_W(p,q)&=A_W(p,q)\star B_W(p,q)\\
	&=\sum_{n=0}^\infty\fft{1}{n!}\left(\fft{i\hbar}{2}\right)^n\sum_{k=0}^\infty(-1)^k\binom{n}{k}(\partial_p^k\partial_q^{n-k}A_W(p,q))(\partial_p^{n-k}\partial_q^kB_W(p,q)),\label{Wigner:product}
\end{split}
\end{equation}
where the $\star$-operation can be written compactly as
\begin{equation}
	\star=\exp[\fft{i\hbar}{2}(\overleftarrow{\partial}_q\overrightarrow{\partial}_p-\overleftarrow{\partial}_p\overrightarrow{\partial}_q)].\label{star:oper}
\end{equation}
Based on this $\star$-operation, we define a $\star$-exponential as
\begin{equation}
	e_\star^{X}\equiv 1+X+\fft{1}{2!}X\star X+\fft{1}{3!}X\star X\star X+\cdots.\label{star:exp}
\end{equation}
The Baker-Campbell-Hausdorff (BCH) formula then naturally follows as
\begin{equation}
\begin{split}
	e_\star^X\star e_\star^Y&=\exp_\star\left[X+Y+\fft12[X,Y]_\star+\fft{1}{12}[X,[X,Y]_\star]_\star-\fft{1}{12}[Y,[X,Y]_\star]_\star\right.\\&\kern3em~\left.-\fft{1}{24}[Y,[X,[X,Y]_\star]_\star]_\star+\mathcal O(\hbar^4)\right].\label{BCH}
\end{split}
\end{equation}
Note that the $\star$-commutator can be computed using the definition (\ref{star:oper}) and its leading order in the semi-classical expansion is given as
\begin{equation}
\begin{split}
	[X,Y]_\star\equiv X\star Y-Y\star X=i\hbar(\partial_qX\partial_pY-\partial_pX\partial_qY)+\mathcal O(\hbar^3).\label{commutator}
\end{split}
\end{equation}

Using the trace formula (\ref{Wigner:trace}) and the Wigner transform of a product of operators (\ref{Wigner:product}), we can derive the expression for the trace of a smooth function of an operator $\hat A$, namely $f(\hat A)$, as
\begin{equation}
\begin{split}
	&\Tr(f(\hat A))\\
	&=\Tr(\sum_{n=0}^\infty\fft{1}{n!}f^{(n)}(A_W(p,q))(\hat A-A_W(p,q))^n)\\
	&=\int\fft{dpdq}{2\pi\hbar}\sum_{n=0}^\infty\fft{1}{n!}f^{(n)}(A_W(p,q))\left[(\hat A-A_W(p,q))^n\right]_W(p,q)\\
	&=\int\fft{dpdq}{2\pi\hbar}f(A_W(p,q))+\sum_{n=2}^\infty\fft{1}{n!}\int\fft{dpdq}{2\pi\hbar}f^{(n)}(A_W(p,q))\left[(\hat A-A_W(p,q))^n\right]_W(p,q).\label{eq:tr:f(A)}
\end{split}
\end{equation}
The above Taylor expansion can be rewritten perturbatively with respect to $\hbar^2$ based on the expansion \cite{Marino:2011eh,Grammaticos:1978tf}
\begin{equation}
	\left[(\hat A-A_W(p,q))^n\right]_W(p,q)=\sum_{j\geq[\fft{n+2}{3}]}\mathcal A_n^{(j)}\hbar^{2j}.\label{def:Anj}
\end{equation}
This tells us that only the quadratic\,($n=2$) and cubic\,($n=3$) orders of the Taylor expansion in (\ref{eq:tr:f(A)}) have non-trivial $\hbar^2$ order coefficients $\mathcal A^{(1)}_n$ as (see (5.6) of \cite{Marino:2011eh})
\begin{subequations}
	\begin{align}
	\mathcal A^{(1)}_2&=\fft14((\partial_q\partial_pA_W)^2-\partial_q^2A_W\partial_p^2A_W),\\
	\mathcal A^{(1)}_3&=\fft14(2\partial_qA_W\partial_pA_W\partial_q\partial_pA_W-(\partial_qA_W)^2\partial_p^2A_W-\partial_q^2A_W(\partial_pA_W)^2).
	\end{align}\label{A:hbar:square}%
\end{subequations}
Hence the semi-classical expansion of (\ref{eq:tr:f(A)}) for small $\hbar$ is given as
\begin{equation}
\begin{split}
	\Tr(f(\hat A))&=\fft{1}{\hbar}\int\fft{dpdq}{2\pi}f(A_W(p,q))\\
	&\quad+\hbar\int\fft{dpdq}{2\pi}\left(\fft12f^{(2)}(A_W(p,q))\mathcal A^{(1)}_2+\fft16f^{(3)}(A_W(p,q))\mathcal A^{(1)}_3\right)+\mathcal O(\hbar^3).\label{eq:tr:f(A):semi}
\end{split}
\end{equation}
%

\subsection{\texorpdfstring{Large-$E$}{Large-E} expansions of integrals}\label{App:Fermi:integrals}
\subsubsection{Building blocks}
\label{App:Fermi:integrals:1}
We start with the building blocks that have been used to compute various integrals in \ref{sec:Fermi:calculation:number}. First we have the large-$E$ expansion of an integral $(l\in\mathbb N)$
\begin{equation}
\begin{split}
	&\int_0^{u_\ast}du\,e^{-l(E-\theta u^2)+\fft{l}{\sqrt2}u}e^{-\fft{A}{\sqrt2}u}\\
	&=\fft12\sqrt{\fft{\pi}{l\theta}}e^{-lE-\fft{\left(l-A\right)^2}{8l\theta}}\left(\mathrm{erfi}(\sqrt{l\theta}(u_\ast+\fft{l-A}{2\sqrt2l\theta}))-\mathrm{erfi}(\sqrt{l\theta}\fft{l-A}{2\sqrt2l\theta})\right)\\
	&=e^{-\fft{A}{\sqrt2}u_\ast}\left(u_\ast+\fft{l-A}{2\sqrt2l\theta}\right)^{-1}\sum_{k=0}^\infty\fft{(2k-1)!!}{(2l\theta)^{k+1}}\left(u_\ast+\fft{l-A}{2\sqrt2l\theta}\right)^{-2k}+\mathcal O(e^{-E})\\
	&=\left(\fft{E}{\theta}\left(1+\fft{1}{8\theta E}\right)\right)^{-\fft12}\sum_{k=0}^\infty\fft{(2k-1)!!}{(2l\theta)^{k+1}}\left(\fft{E}{\theta}\left(1+\fft{1}{8\theta E}\right)\right)^{-k}+\mathcal O(e^{-\sqrt{E/2\theta}}),\label{building:block:0}
\end{split}
\end{equation}
for $A=0$, where the last equation is replaced with $=\mathcal O(e^{-A\sqrt{E/2\theta}})$ for $A>0$. The $u_*$ in the integration range is given in (\ref{eq:u*}). Note that in the second equation we have used the asymptotic expansion of $\mathrm{erfi}(x)\equiv-i\mathrm{erf}(ix)$, namely
\begin{equation}
	\mathrm{erfi}(x)=-i+\fft{e^{x^2}}{\sqrt\pi x}\sum_{k=0}\fft{(2k-1)!!}{(2x^2)^k}.
\end{equation}
Similarly we can derive the following large-$E$ expansions.
\begin{equation}
\begin{split}
	&\int_0^{u_\ast}du\,u^2\,e^{-l(E-\theta u^2)+\fft{l}{\sqrt2}u}e^{-\fft{A}{\sqrt2}u}\\
	&=\fft{l-4\theta}{8l\theta^2}\left(\fft{E}{\theta}\left(1+\fft{1}{8\theta E}\right)\right)^{-\fft12}\sum_{k=0}^\infty\fft{(2k-1)!!}{(2l\theta)^{k+1}}\left(\fft{E}{\theta}\left(1+\fft{1}{8\theta E}\right)\right)^{-k}\\
	&\quad+\fft{1}{2l\theta}\left(\fft{E}{\theta}\left(1+\fft{1}{8\theta E}\right)\right)^\fft12-\fft{1}{2\sqrt2l\theta^2}+\mathcal O(e^{-\sqrt{E/2\theta}}),\label{building:block:2}
\end{split}
\end{equation}
for $A=0$, where the last equation is replaced with $=\mathcal O(e^{-A\sqrt{E/2\theta}})$ for $A>0$, and
\begin{equation}
\begin{split}
	&\int_0^{u_\ast}du\,u\,e^{-l(E-\theta u^2)+\fft{l}{\sqrt2}u}e^{-\fft{A}{\sqrt2}u}\\
	&=\fft{1}{2l\theta}-\fft{1}{2\sqrt2\theta}\left(\fft{E}{\theta}\left(1+\fft{1}{8\theta E}\right)\right)^{-\fft12}\sum_{k=0}^\infty\fft{(2k-1)!!}{(2l\theta)^{k+1}}\left(\fft{E}{\theta}\left(1+\fft{1}{8\theta E}\right)\right)^{-k}\\
	&\quad+\mathcal O(e^{-\sqrt{E/2\theta}}),\label{building:block:1}
\end{split}
\end{equation}
for $A=0$, where the last equation is replaced with $=\mathcal O(e^{-A\sqrt{E/2\theta}})$ for $A>0$.

\subsubsection{Applications}
\label{App:Fermi:integrals:2}
Next we summarize various integrals used to compute the first quantum corrections to the number of states $n^{(1)}(E;\theta)$ in Section~\ref{sec:Fermi:calculation:number}:
\begin{subequations}
\begin{align}
	\int_0^{u_\ast} du\,\fft{\fft12e^{E-\theta u^2}u^2}{\sqrt{(\fft12e^{E-\theta u^2}-\cosh\fft{u}{\sqrt2})^2-1}}&=\fft{1}{3\theta^\fft32}E^\fft32-\fft{3}{16\theta^\fft52}E^\fft12+\fft{1}{6\sqrt2\theta^3}\nn\\
	&\quad-\fft{15+128\theta^2\zeta(2)}{512\theta^\fft72}E^{-\fft12}+\mathcal O(E^{-\fft32}),\label{eq:1:u^2}\\
	\int_0^{u_\ast} du\,\fft{\fft12e^{E-\theta u^2}}{\sqrt{(\fft12e^{E-\theta u^2}-\cosh\fft{u}{\sqrt2})^2-1}}&=\fft{1}{\theta^\fft12}E^\fft12-\fft{1}{16\theta^\fft32}E^{-\fft12}+\mathcal O(E^{-\fft32}),\label{eq:1:u^0}
\end{align}
\end{subequations}
\begin{subequations}
\begin{align}
	\int_0^{u_\ast} du\,\fft{\fft12e^{\fft{u}{\sqrt2}}u^2}{\sqrt{(\fft12e^{E-\theta u^2}-\cosh\fft{u}{\sqrt2})^2-1}}&=\fft{1}{2\sqrt2\theta^2}E-\fft{3}{8\theta^\fft52}E^\fft12+\fft{1}{4\sqrt2\theta^3}\nn\\
	&\quad-\fft{5+32\theta^2\zeta(2)}{128\theta^\fft72}E^{-\fft12}+\mathcal O(E^{-\fft32}),\label{eq:2:u^2}\\
	\int_0^{u_\ast} du\,\fft{\fft12e^{\fft{u}{\sqrt2}}u}{\sqrt{(\fft12e^{E-\theta u^2}-\cosh\fft{u}{\sqrt2})^2-1}}&=\fft{1}{2\sqrt2\theta^\fft32}E^\fft12-\fft{1}{4\theta^2}+\fft{3}{32\sqrt2\theta^\fft52}E^{-\fft12}+\mathcal O(E^{-\fft32}),\label{eq:2:u^1}\\
	\int_0^{u_\ast} du\,\fft{\fft12e^{\fft{u}{\sqrt2}}}{\sqrt{(\fft12e^{E-\theta u^2}-\cosh\fft{u}{\sqrt2})^2-1}}&=\fft{1}{2\sqrt2\theta}-\fft{1}{8\theta^\fft32}E^{-\fft12}+\mathcal O(E^{-\fft32}),\label{eq:2:u^0}
\end{align}
\end{subequations}
\begin{subequations}
\begin{align}
	\int_0^{u_\ast} du\,\fft{\fft12e^{-E+\theta u^2+\sqrt2u}u^2}{\sqrt{(\fft12e^{E-\theta u^2}-\cosh\fft{u}{\sqrt2})^2-1}}&=\fft{1}{2\sqrt2\theta^2}E-\fft{3+4\theta}{8\theta^\fft52}E^\fft12+\fft{1+2\theta}{4\sqrt2\theta^3}\nn\\
	&\quad-\fft{5+12\theta-32\theta^2(1-\zeta(2))}{128\theta^\fft72}E^{-\fft12}+\mathcal O(E^{-\fft32}),\label{eq:3:u^2}\\
	\int_0^{u_\ast} du\,\fft{\fft12e^{-E+\theta u^2+\sqrt2u}u}{\sqrt{(\fft12e^{E-\theta u^2}-\cosh\fft{u}{\sqrt2})^2-1}}&=\fft{1}{2\sqrt2\theta^\fft32}E^\fft12-\fft{1+2\theta}{4\theta^2}+\fft{3+8\theta}{32\sqrt2\theta^\fft52}E^{-\fft12}+\mathcal O(E^{-\fft32}),\label{eq:3:u^1}\\
	\int_0^{u_\ast} du\,\fft{\fft12e^{-E+\theta u^2+\sqrt2u}}{\sqrt{(\fft12e^{E-\theta u^2}-\cosh\fft{u}{\sqrt2})^2-1}}&=\fft{1}{2\sqrt2\theta}-\fft{1+4\theta}{8\theta^\fft32}E^{-\fft12}+\mathcal O(E^{-\fft32}),\label{eq:3:u^0}
\end{align}
\end{subequations}

We explain how to derive (\ref{eq:1:u^2}) then the others can be computed in the same way. To begin with, take the Taylor expansion as
\begin{equation}
	\int_0^{u_\ast(E)} du\,\fft{\fft12e^{E-\theta u^2}u^2}{\sqrt{(\fft12e^{E-\theta u^2}-\cosh\fft{u}{\sqrt2})^2-1}}=\sum_{n=0}^\infty\binom{-1/2}{n}(-1)^nX_n(E).\label{eq:1:u^2:1}
\end{equation}
where we have defined $X_n(E)$ as
\begin{equation}
    X_n(E)\equiv\fft{\fft12e^{E-\theta u^2}u^2}{\left(\fft12e^{E-\theta u^2}-\cosh\fft{u}{\sqrt2}\right)^{2n+1}}.
\end{equation}
We can compute $X_0(E)$ as
\begin{equation}
\begin{split}
	X_0(E)&=\int_0^{u_\ast(E)} du\,\fft{d}{dE}\left[u^2\log(\fft12e^{E-\theta u^2}-\cosh\fft{u}{\sqrt2})\right]\\
	&=\fft{d}{dE}\int_0^{u_\ast(E)}du\,u^2\log(e^{E-\theta u^2}-2\cosh\fft{u}{\sqrt2})\\
	&\quad-\lim_{\epsilon\to0}\fft{1}{\epsilon}\int_{u_\ast(E)}^{u_\ast(E+\epsilon)}du\,u^2\log(e^{E+\epsilon-\theta u^2}-2\cosh\fft{u}{\sqrt2})\\
	&=\fft{d}{dE}\int_0^{u_\ast(E)}du\,\left[u^2(E-\theta u^2)-u^2\sum_{l=1}^\infty\fft{1}{l}\left(2e^{-E+\theta u^2}\cosh\fft{u}{\sqrt2}\right)^l\right]\\
	&\quad-u_\ast'(E)u_\ast(E)^2\log(e^{E-\theta u_\ast(E)^2}-2\cosh\fft{u_\ast(E)}{\sqrt2})\\
	&=\left[\fft{1}{3\theta^\fft32}E^\fft32-\fft{3}{16\theta^\fft52}E^\fft12+\fft{1}{6\sqrt2\theta^3}-\left(\fft{15}{512\theta^\fft72}+\fft{\zeta(2)}{4\theta^\fft52}\right)E^{-\fft12}+\mathcal O(E^{-\fft32})\right]\\
	&\quad~-\log2\left[\fft{1}{2\theta^\fft32}E^\fft12-\fft{1}{2\sqrt2\theta^2}+\fft{3}{32\theta^\fft52}E^{-\fft12}+\mathcal O(E^{-\fft32})\right],\label{eq:X0}
\end{split}
\end{equation}
where we have used the large-$E$ expansion (\ref{building:block:2}). Similarly the other $X_n(E)$ with $n\geq1$ can be computed as
\begin{equation}
\begin{split}
	X_n(E)&=\int_0^{u_\ast(E)} du\,\fft{d}{dE}\left[-\fft{1}{2n}u^2\left(\fft12e^{E-\theta u^2}-\cosh\fft{u}{\sqrt2}\right)^{-2n}\right]\\
	&=\fft{d}{dE}\int_0^{u_\ast(E)}du\,\left[-\fft{1}{2n}u^2\left(\fft12e^{E-\theta u^2}-\cosh\fft{u}{\sqrt2}\right)^{-2n}\right]\\
	&\quad-\lim_{\epsilon\to0}\fft{1}{\epsilon}\int_{u_\ast(E)}^{u_\ast(E+\epsilon)}du\,\left[-\fft{1}{2n}u^2\left(\fft12e^{E+\epsilon-\theta u^2}-\cosh\fft{u}{\sqrt2}\right)^{-2n}\right]\\
	&=-u_\ast'(E)\left[-\fft{1}{2n}u_\ast(E)^2\left(\fft12e^{E-\theta u_\ast(E)^2}-\cosh\fft{u_\ast(E)}{\sqrt2}\right)^{-2n}\right]+\mathcal O(e^{-\sqrt{E/2\theta}})\\
	&=\fft{1}{2n}\left[\fft{1}{2\theta^\fft32}E^\fft12-\fft{1}{2\sqrt2\theta^2}+\fft{3}{32\theta^\fft52}E^{-\fft12}+\mathcal O(E^{-\fft32})\right].\label{eq:Xn}
\end{split}
\end{equation}
Substituting (\ref{eq:X0}) and (\ref{eq:Xn}) into (\ref{eq:1:u^2:1}) and using the identity
\begin{equation}
	\sum_{n=1}^\infty\binom{-1/2}{n}\fft{(-1)^n}{2n}=\log 2,
\end{equation}
we get (\ref{eq:1:u^2}).

\subsection{Bounds in \texorpdfstring{large-$\mu$}{large-mu} expansions}\label{App:Fermi:grand:potential}
Here we argue that the integrals in (\ref{eq:J}) are of order $\mathcal O(\mu^0)$ at most. The first integral in the second line of (\ref{eq:J}) is bounded as
\begin{equation}
\begin{split}
	\left|\int_0^{E_0}dE\,\fft{\alpha E^\fft32+\beta E^\fft12+\delta_0 E^{-\fft12}}{1+e^{E-\mu}}\right|&\leq\int_0^{E_0}dE\,\left(|\alpha|E^\fft32+|\beta|E^\fft12+|\delta_0|E^{-\fft12}\right)\\
	&=\fft25|\alpha|E_0^\fft52+\fft23|\beta|E_0^\fft32+2|\delta_0|E_0^\fft12,
\end{split}
\end{equation}
which is finite in the large-$\mu$ regime. So it is of order $\mathcal O(\mu^0)$. 

To estimate the second integral in the second line of (\ref{eq:J}), we split it into two parts using (\ref{ntilde:property}) as
\begin{equation}
\begin{split}
	\int_{E_0}^\infty dE\,\tilde n(E;\theta,\hbar)&=\int_{E_0}^{E_\ast(\theta,\hbar)}dE\,\tilde n(E;\theta,\hbar)\\
	&\quad+\int_{E_\ast(\theta,\hbar)}^\infty dE\,\left(\sum_{n=1}^\infty\delta_nE^{-(n+\fft12)}+\mathcal O(e^{-\mathcal B(\theta,\hbar)\sqrt E})\right).
\end{split}
\end{equation}
The first part is independent of $\mu$ and therefore obviously $\mathcal O(\mu^0)$.
The second part converges in the large-$\mu$ regime so it is of order $\mathcal O(\mu^0)$ too. 

Lastly, to estimate the third integral of (\ref{eq:J}), we split the integration range into $[E_0,E_*(\theta,\hbar)]$ and $(E_*(\theta,\hbar),\infty)$ as above. The first part is then bounded as
\begin{equation}
	\left|\int_{E_0}^{E_\ast(\theta,\hbar)}dE\,\fft{\tilde n(E;\theta,\hbar)}{1+e^{\mu-E}}\right|\leq\max_{E\in[E_0,E_\ast(\theta,\hbar)]}|\tilde n(E;\theta,\hbar)|\times\log(\fft{1+e^{E_\ast(\theta,\hbar)-\mu}}{1+e^{E_0-\mu}}).
\end{equation}
Since $\max_{E\in[E_0,E_\ast(\theta,\hbar)]}|\tilde n(E;\theta,\hbar)|$ is finite and independent of $\mu$, the RHS and therefore the LHS is exponentially suppressed as $\mathcal O(e^{-\mu})$. The second part is bounded as
\begin{equation}
\begin{split}
	\left|\int_{E_\ast(\theta,\hbar)}^\infty dE\,\fft{\tilde n(E;\theta,\hbar)}{1+e^{\mu-E}}\right|&\leq\int_{E_\ast(\theta,\hbar)}^\infty dE\,\fft{\sum_{n=1}^\infty|\delta_n|E^{-(n+\fft12)}+\mathcal A(\theta,\hbar)e^{-\mathcal B(\theta,\hbar)\sqrt E}}{1+e^{\mu-E}}\\
	&\leq\sum_{n=1}^\infty\fft{|\delta_n|}{n-\fft12}(E_\ast(\theta,\hbar))^{-n+\fft12}+2\mathcal A(\theta,\hbar)\int_0^\infty dx\,xe^{-\mathcal B(\theta,\hbar)x},\\
	&\leq\sum_{n=1}^\infty\fft{|\delta_n|}{n-\fft12}(E_\ast(\theta,\hbar))^{-n+\fft12}+\fft{2\mathcal A(\theta,\hbar)}{\mathcal B(\theta,\hbar)^2},\label{subtle}
\end{split}
\end{equation}
where we have used (\ref{ntilde:property}) in the first inequality. Since we have assumed the absolute convergence of the series $\sum_{n=1}^\infty|\delta_n|(E_*(\theta,\hbar))^{-(n+\fft12)}$ below (\ref{eq:n(E):largeE}), the RHS of (\ref{subtle}) becomes finite in the large-$\mu$ regime. Hence we conclude that the second part is also of order $\mathcal O(\mu^0)$.

\section{Obtaining the ideal Fermi-gas partition function}\label{App:Fermi:setup}

Here we review how the $S^3$ partition function of the GT theory (\ref{eq:Z2}) can be rewritten as a partition function of an ideal Fermi-gas model \cite{Marino:2011eh}. To begin with, applying the identity (see (4.2) of \cite{Kapustin:2010xq} for example),
\begin{equation}
	\fft{\prod_{i<j}\sinh(x_{ij})\sinh(y_{ij})}{\prod_{i,j}\cosh(x_i-y_j)}=\sum_{\sigma\in S_N}(-1)^{\epsilon(\sigma)}\prod_i\fft{1}{\cosh(x_i-y_{\sigma(i)})},
\end{equation}
we can rewrite (\ref{eq:Z2}) as
\begin{equation}
	Z=\fft{1}{(N!)^2}\sum_{\sigma,\tilde\sigma\in S_n}(-1)^{\epsilon(\sigma)+\epsilon(\tilde\sigma)}\int\fft{d^N\mu}{(4\pi)^N}\fft{d^N\nu}{(4\pi)^N}\prod_{i=1}^N\fft{\exp[\fft{i}{4\pi}(k_1\mu_i^2+k_2\nu_i^2)]}{\cosh(\fft{\mu_i-\nu_{\sigma(i)}}{2})\cosh(\fft{\mu_i-\nu_{\tilde\sigma(i)}}{2})}.
\label{eq:Z2inter}
\end{equation}
Renaming $\nu_i$ as $\nu_i\to\nu_{\tilde\sigma^{-1}(i)}$ then redefining the permutations as 
\begin{equation}
	\sigma\to\tilde\sigma\circ\sigma,\qquad\tilde\sigma\to\tilde\sigma,
\end{equation}
this can be simplified as
\begin{equation}
	Z=\fft{1}{N!}\sum_{\sigma\in S_n}(-1)^{\epsilon(\sigma)}\int\fft{d^N\mu}{(4\pi)^N}\fft{d^N\nu}{(4\pi)^N}\prod_{i=1}^N\fft{\exp[\fft{i}{4\pi}(k_1\mu_i^2+k_2\nu_i^2)]}{\cosh(\fft{\mu_i-\nu_i}{2})\cosh(\fft{\mu_i-\nu_{\sigma(i)}}{2})}.
\end{equation}
Next, replacing hyperbolic cosine functions in the denominator with integrals using
\begin{equation}
	\fft{1}{\cosh(\eta/2)}=\int\fft{d\rho}{2\pi}\fft{e^{\fft{i}{2\pi}\eta\rho}}{\cosh(\rho/2)},
\end{equation}
we have
\begin{equation}
\begin{split}
	Z&=\fft{1}{N!}\sum_{\sigma\in S_n}(-1)^{\epsilon(\sigma)}\int\fft{d^Nx}{(2\pi)^N}\fft{d^Ny}{(2\pi)^N}\int\fft{d^N\mu}{(4\pi)^N}\fft{d^N\nu}{(4\pi)^N}\\
	&\kern9em\prod_{i=1}^N\fft{\exp[\fft{i}{2\pi}((\mu_i-\nu_i)x_i+(\mu_i-\nu_{\sigma(i)})y_i)+\fft{i}{4\pi}(k_1\mu_i^2+k_2\nu_i^2)]}{\cosh(x_i/2)\cosh(y_i/2)}.
\end{split}
\end{equation}
Then the Gaussian integral over $\mu_i$ and $\nu_i$ gives
\begin{equation}
\begin{split}
	Z&=\fft{1}{N!}\sum_{\sigma\in S_n}(-1)^{\epsilon(\sigma)}\int\fft{d^Nx}{(4\pi)^N}\fft{d^Ny}{(4\pi)^N}(\fft{i}{k_1})^{\fft{N}{2}}(\fft{i}{k_2})^{\fft{N}{2}}\\
	&\kern12em\prod_{i=1}^N\fft{\exp[-\fft{i}{4\pi}(\fft{1}{k_1}+\fft{1}{k_2})(x_i^2+y_i^2)-\fft{i}{2\pi}x_i(\fft{1}{k_1}y_i+\fft{1}{k_2}y_{\sigma^{-1}(i)})]}{\cosh(x_i/2)\cosh(y_i/2)}\\
	&=\fft{1}{N!}\sum_{\sigma\in S_n}(-1)^{\epsilon(\sigma)}\int d^Nx\,d^Ny\\
	&\kern7.5em\prod_{i=1}^N\fft{(1+\fft14\Delta^2\theta^2)^\fft12}{8\pi|\Delta|}\fft{\exp[-\fft{\theta}{2}(x_i^2+y_i^2)-\fft{\theta}{2}(x_i+x_{\sigma(i)})y_i+\fft{i}{\Delta}(x_i-x_{\sigma(i)})y_i]}{\cosh(x_i/2)\cosh(y_i/2)},\label{fermi:gas:ptn:fct:explicit}
\end{split}
\end{equation}
where we have chosen principal values for fractional exponents and defined $\theta$ and $k$ as
\begin{equation}
	2\pi i\theta=-\fft{1}{k_1}-\fft{1}{k_2},\qquad \fft{4\pi}{\Delta}=\fft{1}{k_2}-\fft{1}{k_1}.
\end{equation}
The final result (\ref{fermi:gas:ptn:fct:explicit}) can be written compactly as (\ref{fermi:gas:ptn:fct}).

\bibliographystyle{JHEP}
\bibliography{logN}

\end{document}